\documentclass[draftclsnofoot,onecolumn]{IEEEtran}
\usepackage{graphicx,subfigure,psfrag,epsfig,epsf,latexsym,hhline,amsmath,amssymb,amsthm,multirow}

\begin{document}

\title{Multilevel Coding Schemes for Compute-and-Forward with Flexible Decoding\thanks{This work was supported by the National Science Foundation under Grant CCF 0729210. Parts of this work have been published at the 2011 IEEE International Symposium on Information Theory.}}
\author{Brett Hern and Krishna Narayanan\\
Department of Electrical and Computer Engineering \\
Texas A\&M University\\
College Station\\
TX 77843, U.S.A}
\maketitle

\begin{abstract}
We consider the design of coding schemes for the wireless two-way relaying channel when there is no channel state information at the transmitter. In the spirit of the compute and forward paradigm, we present a multilevel coding scheme that permits computation (or, decoding) of a class of functions at the relay. The function to be computed (or, decoded) is then chosen depending on the channel realization. We define such a class of functions which can be decoded at the relay using the proposed coding scheme and derive rates that are universally achievable over a set of channel gains when this class of functions is used at the relay. We develop our framework with general modulation formats in mind, but numerical results are presented for the case where each node transmits using the QPSK constellation. Numerical results with QPSK show that the flexibility afforded by our proposed scheme results in substantially higher rates than those achievable by always using a fixed function or by adapting the function at the relay but coding over GF(4).
\end{abstract}


\section{Introduction}
Physical layer network coding (PLNC) or Compute and Forward is a new paradigm in wireless networks where each relay in a network decodes a function of the transmitted messages and broadcasts the value of this function to the other nodes in the network. This has been shown to provide significant increase in achievable rates for some networking problems \cite{popovski2009coded}, \cite{DBLP:journals/corr/abs-0805-0012}, \cite{nazer2007lattice}. For a recent and approachable tutorial/survey of the key ideas behind PLNC with reliable decoding, we refer readers to \cite{nazer2011reliable}. For another broad tutorial/survey of PLNC results with an eye to practical implementation, we refer readers to \cite{liew2011physical}.

\begin{figure}
    \centering
    \includegraphics[width=3.5in]{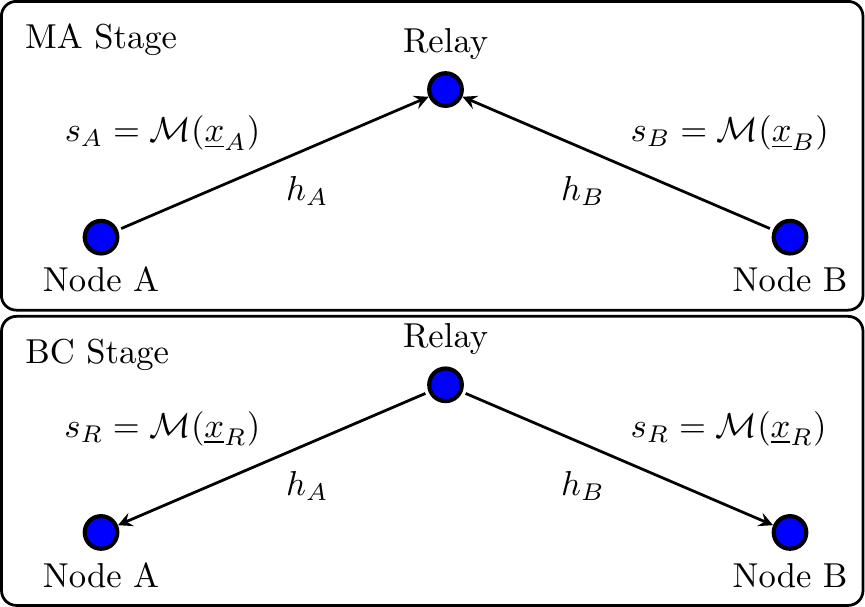}
    \caption{System model showing a of two-way relay channel with PLNC.}
    \label{fig:RelayChannel}
\end{figure}

An example of such a problem where compute and forward has been shown to be effective is the two-way relaying system shown in Fig. \ref{fig:RelayChannel}. Here, node $A$ has data to send to node $B$ and vice versa. The relay $R$ is included to assist in this communication, and it is assumed that there is no direct link between nodes $A$ and $B$. Near optimal coding schemes have been designed to maximize the exchange rate for the case where there is no fading in the channel in \cite{DBLP:journals/corr/abs-0805-0012}, \cite{nazer2007lattice}, \cite{nam5capacity}. Building on results from \cite{erez2004achieving}, these authors derive an upper bound on the capacity of $\frac{1}{2}\log(1+snr)$ and show that with lattice coding and lattice decoding a rate of $\frac{1}{2}\log(\frac{1}{2}+snr)$ is achievable. This problem has also been studied for case where there is fading in the channel, but each node perfectly knows the fading coefficients for each network link in \cite{wilson2009power}. It has been shown that near-optimal performance can be obtained at high signal-to-noise ratio (SNR) if each transmitter inverts its channel prior to transmission. The authors in \cite{song2010list} apply lattices with list decoding to the two way relaying problem with a direct link between nodes A and B. Finally, compute and forward schemes for multiple input multiple output channels have been considered in \cite{zhan2009mimo}.

In this paper, the complex channel coefficients $h_A$ and $h_B$ are assumed to be perfectly estimated at each receiver but unknown to each transmitter. For this scenario, the authors in \cite{koike2009optimized} introduce a scheme called denoise-and-forward which uses channel dependent denoising functions at the relay to minimize the symbol error probability. The relay chooses denoising functions so that the distance profile for constellation points with different labels is optimized. This improves the symbol error rate for transmissions between nodes A and B, however,
denoising is performed purely at the symbol level. There is no natural extension to include error correction at the relay.

Recently, a scheme called compute-and-forward, which allows both adaptation of decoding functions and error correction at the relay has been presented in \cite{nazer2009compute}. In this scheme, the relay decodes an integer combination of the transmitted codewords, where the integer combination is adapted according to the channel gains. They show that such a scheme can be implemented using nested lattice codes to take advantage of the duality between modulus arithmetic in prime order fields and the modular operations of lattice decoding. Their scheme requires the construction of infinite dimensional lattice codes which is not practical. The results in \cite{nazer2009compute} are extended in a remarkable way in \cite{feng2010algebraic}, where an algebraic framework is provided to design lattices over principal ideal domains. However, their proposed coding scheme is also based on large dimensional lattice codes.

In this paper, we propose a compute and forward scheme based on multilevel coding (MLC). Unlike the coding schemes in \cite{nazer2009compute}, \cite{feng2010algebraic},  our proposed scheme does not result in a lattice code and uses only linear codes over small prime fields (for example, binary linear codes), and can therefore be implemented with lower encoding and decoding complexity. Yet, it facilitates error correction for a larger class of decoding functions than those proposed in \cite{nazer2009compute}. This is because the class of functions for our scheme is derived from the large set of non-singular square matrices over $\mathbb{F}_p$ in place of the set of non-zero elements in large prime order fields. To the best of our knowledge, such an idea of using multilevel coding and exploiting the linearity over the prime field to adaptively decode linear functions of transmitted codewords is new. Another important contribution in this paper is that our proof for the achievability of rates with the proposed multilevel coding scheme requires a non-trivial extension of the proof of achievability of rates for multilevel coding for the point to point case.


This paper is organized as follows. The key elements of the problem are outlined in Section \ref{sec:ProblemDescription}. Our proposed solution is detailed in Section \ref{sec:ProposedScheme}. An achievable rate for the proposed scheme during the MA stage is given in Section \ref{sec:InfoAnalysis}. These rates are numerically determined for an example where nodes A and B transmit using a QPSK constellation in Section \ref{sec:NumericalResults}. Simulation results for a regular LDPC code are shown to corroborate the information theoretic results. Key results are reiterated in Section \ref{sec:ConcludingRemarks}.

Throughout this paper we will use the following naming conventions. Vectors or sequences will be denoted by underlined variables such as $\underline{x}$. Random variables will be denoted by upper case variables such as $X$, while their outcomes will be represented by lowercase variables.  Matrices will be represented by capital boldface letters such as $\mathbf{X}$. Subsets will be denoted by capital scripted letters such as $\mathcal{X}$. If a variable is associated with a specific node, this will be indicated by a subscripted capital letter like $x_A$. MLC sometimes requires us to split a data sequence into subsequences for parallel encoding and transmission over separate bit levels. Variables associated with a specific bit level will be indicated by a superscript like $x^k$. A specific element of a vector or sequence will be referred to by an index in brackets like $\underline{x}[n]$.

\section{Problem Description} \label{sec:ProblemDescription}
Each node in the relay network is assumed to be half-duplex, so communication is split into two stages, a multiple access (MA) stage and a broadcast (BC) stage. We assume perfect synchronization between the transmitters and mainly focus on the MA stage in this paper.

\subsection{Multiple Access Stage}
Nodes A and B each encode their binary messages $\underline{u}_A$ and $\underline{u}_B$ into codewords $\underline{v}_A\in\mathcal{C}_A$ and $\underline{v}_B\in\mathcal{C}_B$ where $\mathcal{C}_A$ and $\mathcal{C}_B$ are the codebooks used at the nodes $A$ and $B$ respectively. These codewords are mapped to sequences of symbols $\underline{s}_A,\underline{s}_B\in\mathcal{Q}^N$ with $|\mathcal{Q}|=2^\ell$. The relay receives noisy observations of the sum of these symbol sequences according to
\begin{equation}
\underline{y}_R = h_A\underline{s}_A+h_B\underline{s}_B+\underline{w}_R
\end{equation}
where $h_A$ and $h_B$ are complex fading coefficients, and $\underline{w}_R$ is complex additive white Gaussian noise (AWGN). This induces an effective constellation $\mathcal{Q}_R$ at the relay defined by
\begin{equation} \label{eq:RelayConst}
\mathcal{Q}_R = \{q_R\in\mathbb{C} | q_R = h_Aq_A+h_Bq_B , ~ q_A,q_B\in\mathcal{Q} \}.
\end{equation}

\subsection{Adaptive Decoding at the Relay}
The main idea proposed in this paper is the construction of a coding scheme such that the relay can reliably decode some function of $\underline{v}_A$ and $\underline{v}_B$ for a desired set of channel conditions $\mathcal{H}\subset\mathbb{C}^2$. Specifically, we jointly design codes $\mathcal{C}_A$ and $\mathcal{C}_B$ and a set of decoding functions $\mathcal{F}$ such that, for any $(h_A,h_B)\in\mathcal{H}$, there exists $f\in\mathcal{F}$ such that the relay can reliably decode $f(\underline{v}_A,\underline{v}_B)$ from $\underline{y}_R$. We require that node A (B) must be able to unambiguously decode $\underline{v}_B~(\underline{v}_A)$ from the output of $f(\underline{v}_A,\underline{v}_B)$ with its knowledge of $\underline{v}_A~(\underline{v}_B)$. For a given $f\in\mathcal{F}$, we will define an induced codebook at the relay as the codebook corresponding to $f$ i.e.
\begin{equation}
\mathcal{C}_{f,R} = \{f(\underline{v}_A,\underline{v}_B)| \underline{v}_A \in \mathcal{C}_A,~ \underline{v}_B \in \mathcal{C}_B \}.
\end{equation}
It is important to understand the structure of $\mathcal{C}_{f,R}$ since the probability of error in decoding $f(\underline{v}_A,\underline{v}_B)$ from $\underline{y}_R$ depends on $h_A$, $h_B$, and $\mathcal{C}_{f,R}$. The main advantage of our proposed scheme is that it guarantees that choosing one codebook $\mathcal{C}_A$ and $\mathcal{C}_B$ at the transmitter can result in a good induced codebook
$\mathcal{C}_{f,R}$ for a class of functions $\mathcal{F}$. More specifically, it guarantees $\mathcal{C}_{f,R}$ is a member of the ensemble of random coset codes which is an optimal ensemble for achieving the uniform input information rate for the equivalent channel between $f(\underline{v}_A,\underline{v}_B)$ and $\underline{y}_R$ for all $f \in \mathcal{F}$. We restrict our attention to classes of functions $\mathcal{F}$ which are applied componentwise at the relay.

The broadcast stage is fairly standard and is identical to that considered in \cite{DBLP:journals/corr/abs-0805-0012}, \cite{nazer2007lattice}.


\section{Proposed Scheme} \label{sec:ProposedScheme}

\subsection{Multilevel Encoder}
\begin{figure}
    \centering
    \includegraphics[width=3.5in]{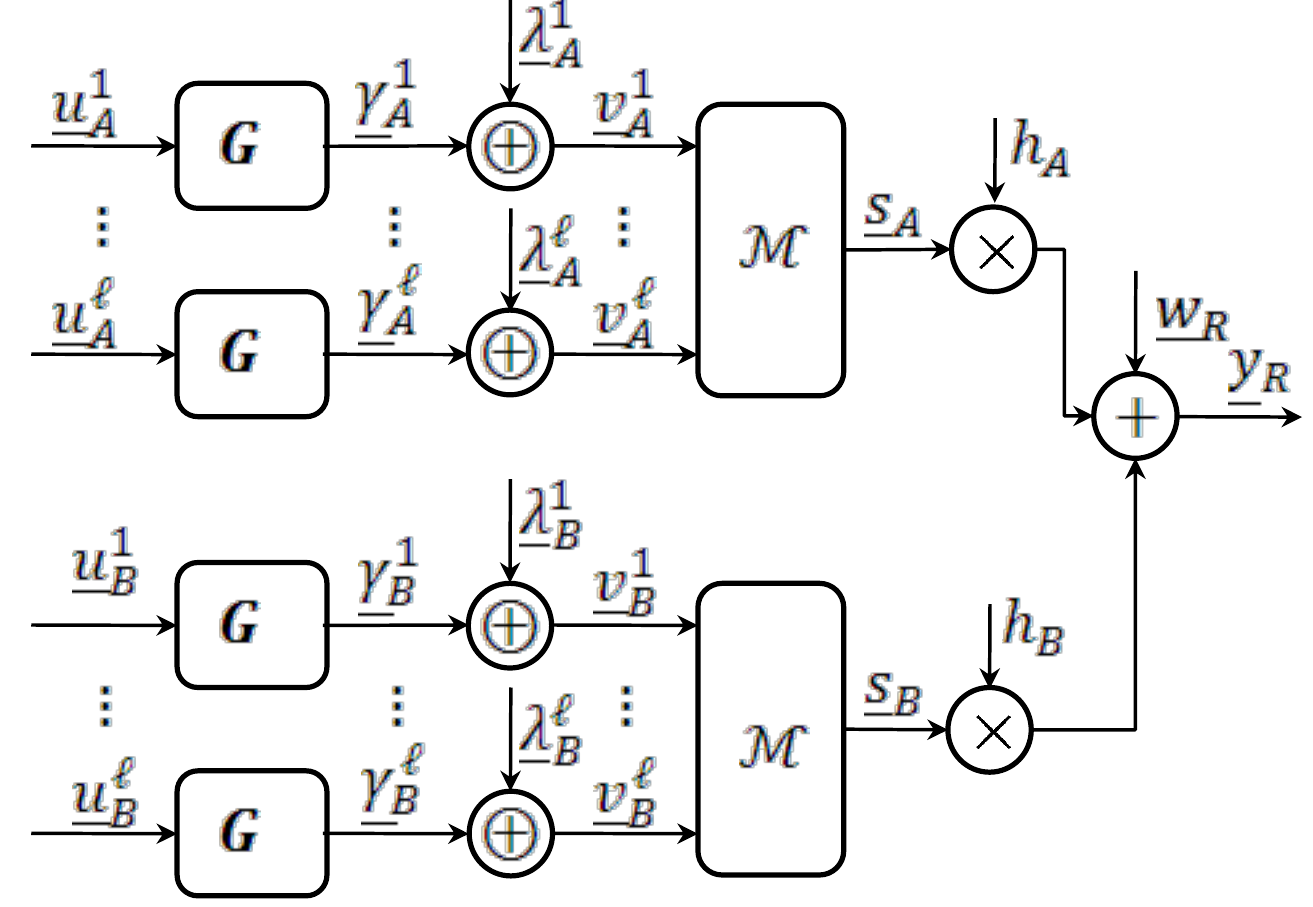}
    \caption{Block Diagram of MLC Coset Encoders for MA Stage.}

    \label{fig:MLCCosetEnc}
\end{figure}

The system model for the multilevel encoder for nodes A and B and the channel model for the MA stage is shown in Fig. \ref{fig:MLCCosetEnc}. The encoder at nodes A and B uses MLC with a different coset of the \emph{same linear} code $\mathcal{C}$ used at each bit level. For a detailed description of MLC and achievable rates for the point to point channel see \cite{wachsmann1999multilevel}.

The encoder is described as it pertains to node A to simplify notation. First, the message $\underline{u}_A$ is split into sub-vectors $\underline{u}_A^1,...,\underline{u}_A^\ell$ which form rows of an $\ell\times K$ matrix
\begin{equation}
\mathbf{U}_A = \left[\begin{array}{c} \underline{u}_A^1 \\ \vdots \\ \underline{u}_A^\ell \end{array} \right].
\end{equation}
Each $\underline{u}_A^k,~\{1,...,\ell\}$ is encoded with a linear code $\mathcal{C}$ with generator matrix $\mathbf{G}$ to get codewords $\underline{\gamma}_A^1,...,\underline{\gamma}_A^\ell$. These codewords from the rows of an $\ell\times N$ matrix
\begin{equation}
\mathbf{\Gamma}_A = \mathbf{U}_A \mathbf{G} = \left[\begin{array}{c} \underline{\gamma}_A^1 \\ \vdots \\ \underline{\gamma}_A^\ell \end{array} \right].
\end{equation}
Finally, a random binary vector $\underline{\lambda}_A^k$ is added to each $\underline{\gamma}_A^k$. Each $\underline{\lambda}_A^k$ can be thought of as coset leaders of a random coset of the original linear code. We obtain a codeword of a random coset given by $\underline{v}_A^k=\underline{\lambda}_A^k\oplus\underline{\gamma}_A^k,~k\in\{1,..,\ell\}$. The random coset leaders form an $\ell\times N$ matrix
\begin{equation}
\mathbf{\Lambda}_A = \left[\begin{array}{c} \underline{\lambda}_A^1 \\ \vdots \\ \underline{\lambda}_A^\ell \end{array} \right].
\end{equation}
The resulting coset codewords $\underline{v}_A^k$ form the rows of a binary $\ell\times N$ matrix $\mathbf{X}_A$ given by
\begin{equation} \label{eq:xAdef}
\mathbf{X}_A = \mathbf{U}_A \mathbf{G} \oplus \mathbf{\Lambda}_A = \left[\begin{array}{c} \underline{v}_A^1 \\ \vdots \\ \underline{v}_A^\ell \end{array} \right] = \left[\underline{x}_A[1],...,\underline{x}_A[N]\right].
\end{equation}
Thus each code $\mathcal{C}_A^k,~k\in\{1,...,\ell\}$ will be a different coset of $\mathcal{C}$. The $k_{th}$ row $\underline{v}_A^k$ of $\mathbf{X}_A$ is then a codeword of $\mathcal{C}_A^k$. We use the two variables $\underline{x}_A[n]$ and $\underline{v}_A^k$ to refer to the $n_{th}$ column and $k_{th}$ row of $\mathbf{X}_A$ respectively because it will simplify our notation later. It should be mentioned here that much of the intuition about the main result in the paper is best obtained by ignoring the fact that cosets are used at each layer and simply considering the use of identical linear codes at each level in the MLC scheme. The coset matrix $\mathbf{\Lambda}_A$ is included to symmetrize the effective channel at the relay (i.e. $\mathbf{\Lambda}_A$ is necessary for the proofs to be correct).

\begin{figure}
    \centering
    \includegraphics[width=3.5in]{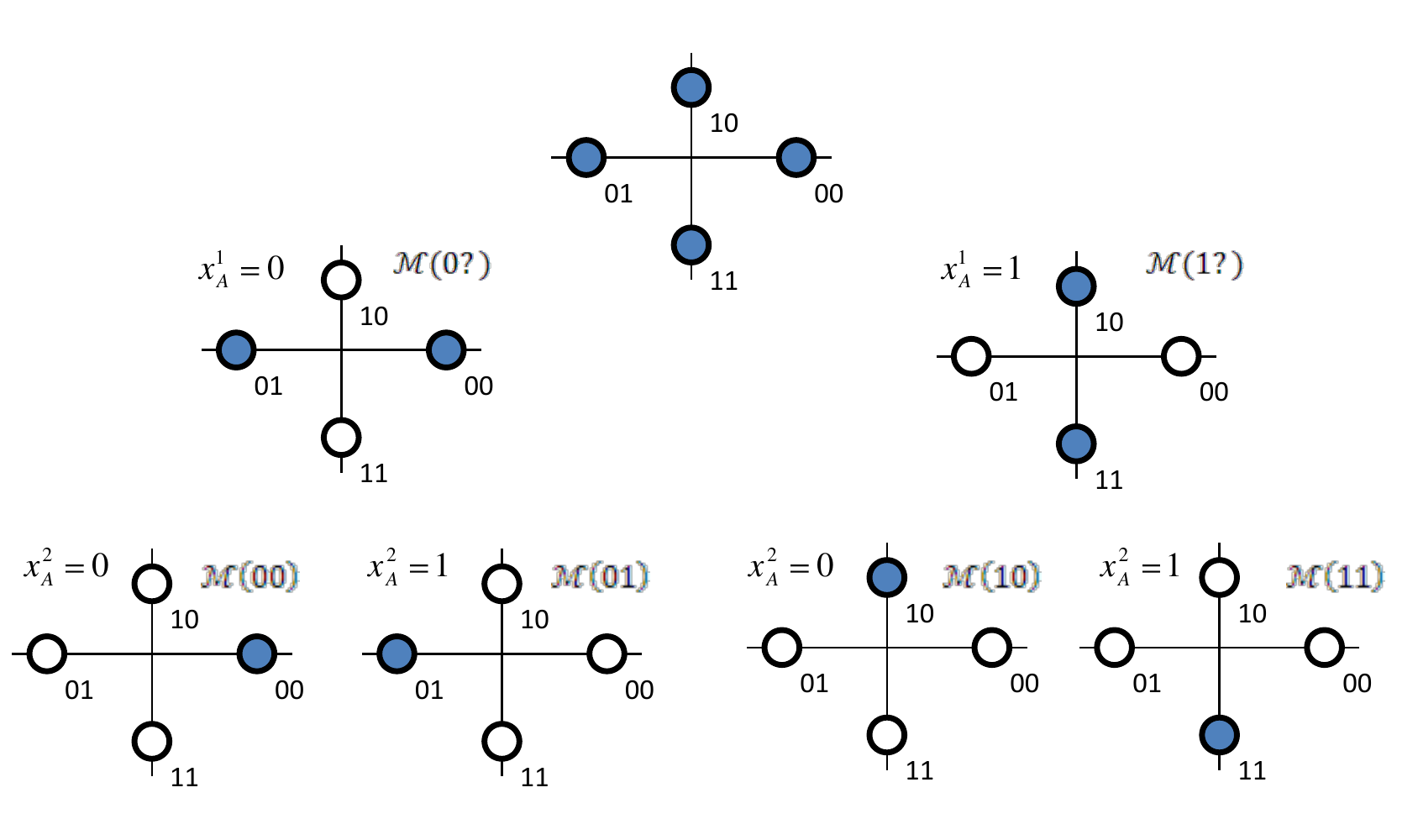}
    \caption{MLC address mapping example using QPSK.}

    \label{fig:MLCEncEx}
\end{figure}

The $n_{th}$ binary address vector $\underline{x}_A[n]\in\mathbb{F}_2^\ell$ maps to a symbol $\underline{s}_A[n]\in\mathcal{Q}$ through the use of a symbol mapping function $\mathcal{M}:\mathbb{F}_2^\ell\rightarrow\mathcal{Q}$. An example of such a mapping function is given in Fig. \ref{fig:MLCEncEx} where $\mathcal{Q}$ is the QPSK constellation. As shown, the mapping function is usually derived by partitioning the set of signaling points in $\mathcal{Q}$ into equal sized subsets. Let $\mathcal{S}\subseteq\{1,...,\ell\}$ be the subset of elements of $\underline{x}_A$ which are fixed. Then we define the output of $\mathcal{M}$ with these input bits as a subset of points from $\mathcal{Q}$ according to
\begin{align} \label{eq:MapDef}
\mathcal{M}(\{x_A^k|k\in\mathcal{S}\})=\{q\in\mathcal{Q}&|q=\mathcal{M}(\{x_A^k|k\in\mathcal{S}\},\{b^i|i\in\overline{\mathcal{S}}\}),~ b^i\in\{0,1\}\}.
\end{align}
This means that the returned subset of constellation points is the subset whose address vectors are equal to the known bits for all indexes, $\mathcal{S}$. The output of $\mathcal{M}(\{x_A^k|k\in\mathcal{S}\})$ is $2^{\ell-|\mathcal{S}|}$ constellation points.

In this example of Fig. \ref{fig:MLCEncEx} if $\mathcal{S}=\{2\}\subseteq\{1,2\}$ and $x_A^2=1$, then
\[
\mathcal{M}(\{x_A^k|k\in\mathcal{S}\}) = \mathcal{M}(?1) = \{\mathcal{M}(01),\mathcal{M}(11)\} = \{-1,-j\}.
\]
Here, $|\mathcal{S}|=1$ and $\ell=2$. Therefore $\mathcal{M}$ returns $2^{2-1}=2$ constellation points.

\subsection{Adaptive Decoding at the Relay}
As mentioned previously, the goal of the proposed scheme is to allow the relay to decode a function of the transmitted codewords. Similar to the compute and forward scheme, our scheme utilizes the linearity of the base code $\mathcal{C}$ and the fact that the relay knows $\mathbf{\Lambda}_A$ and $\mathbf{\Lambda}_B$. If nodes A and B encode their messages as described, the set of decoding functions $\mathcal{F}$ which the relay can use for decoding is defined as follows.

Define $\mathcal{D}$ as the set of $\ell\times\ell$ binary matrices which are invertible over $\mathbb{F}_2$. The set of functions we consider is given by
\begin{align} \label{eq:FSetDef}
\mathcal{F} &= \{f:\mathbb{F}_{2}^\ell\times\mathbb{F}_{2}^\ell\rightarrow\mathbb{F}_{2}^\ell | f(\underline{x}_A,\underline{x}_B) = [\mathbf{D}_A \mathbf{D}_B] \left[\begin{array}{c}\underline{x}_A \\ \underline{x}_B \end{array}\right], ~ \mathbf{D}_A,\mathbf{D}_B\in\mathcal{D} \}.
\end{align}
Therefore a given $f\in\mathcal{F}$ is defined by some $\mathbf{D}_A,\mathbf{D}_B\in\mathcal{D}$ from which the relay should attempt to decode a matrix $\mathbf{X}_{f,R}$ given by
\begin{align}
\mathbf{X}_{f,R} = [\mathbf{D}_A \mathbf{D}_B]
\left[ \begin{array}{c} \mathbf{X}_A \\ \mathbf{X}_B \end{array} \right].
\end{align}

Due to the linearity of $[\mathbf{D}_A,\mathbf{D}_B]$ and $\mathbf{G}$, we can express the desired matrix $\mathbf{X}_{f,R}$ as
\begin{align} \label{eq:XrEquiv}
\mathbf{X}_{f,R} &= [\mathbf{D}_A \mathbf{D}_B] \left[ \begin{array}{c} \mathbf{X}_{A} \\  \mathbf{X}_{B} \end{array} \right] = [\mathbf{D}_A \mathbf{D}_B] \left[ \begin{array}{c} \mathbf{U}_{A} \mathbf{G} \oplus \mathbf{\Lambda}_A \\  \mathbf{U}_{B} \mathbf{G} \oplus \mathbf{\Lambda}_B \end{array} \right] \nonumber \\
&= [\mathbf{D}_A \mathbf{D}_B] \left[ \begin{array}{c} \mathbf{U}_{A} \mathbf{G} \\  \mathbf{U}_{B} \mathbf{G} \end{array} \right] \oplus [\mathbf{D}_A \mathbf{D}_B] \left[ \begin{array}{c} \mathbf{\Lambda}_A \\  \mathbf{\Lambda}_B \end{array} \right] \nonumber  \\
&= [\mathbf{D}_A \mathbf{D}_B] \left[ \begin{array}{c} \mathbf{U}_{A} \\  \mathbf{U}_{B} \end{array} \right] \mathbf{G} \oplus [\mathbf{D}_A \mathbf{D}_B] \left[ \begin{array}{c} \mathbf{\Lambda}_A \\  \mathbf{\Lambda}_B \end{array} \right] \nonumber  \\
&= \mathbf{U}_{f,R} \mathbf{G} \oplus \mathbf{\Lambda}_{f,R}.
\end{align}
Here, we see that the matrix $\mathbf{X}_{f,R}$ can be written in terms of an effective message $\mathbf{U}_{f,R}$ and coset matrix $\mathbf{\Lambda}_{f,R}$ which can be computed separately based on $f$. Thus the rows of $\mathbf{X}_{f,R}$ are codewords from a different coset code of $\mathcal{C}$. Note that $f$ is applied elementwise to the sequences $\underline{s}_A$ and $\underline{s}_B$.

\begin{figure*}
    \centering
    \includegraphics[width=0.8\textwidth]{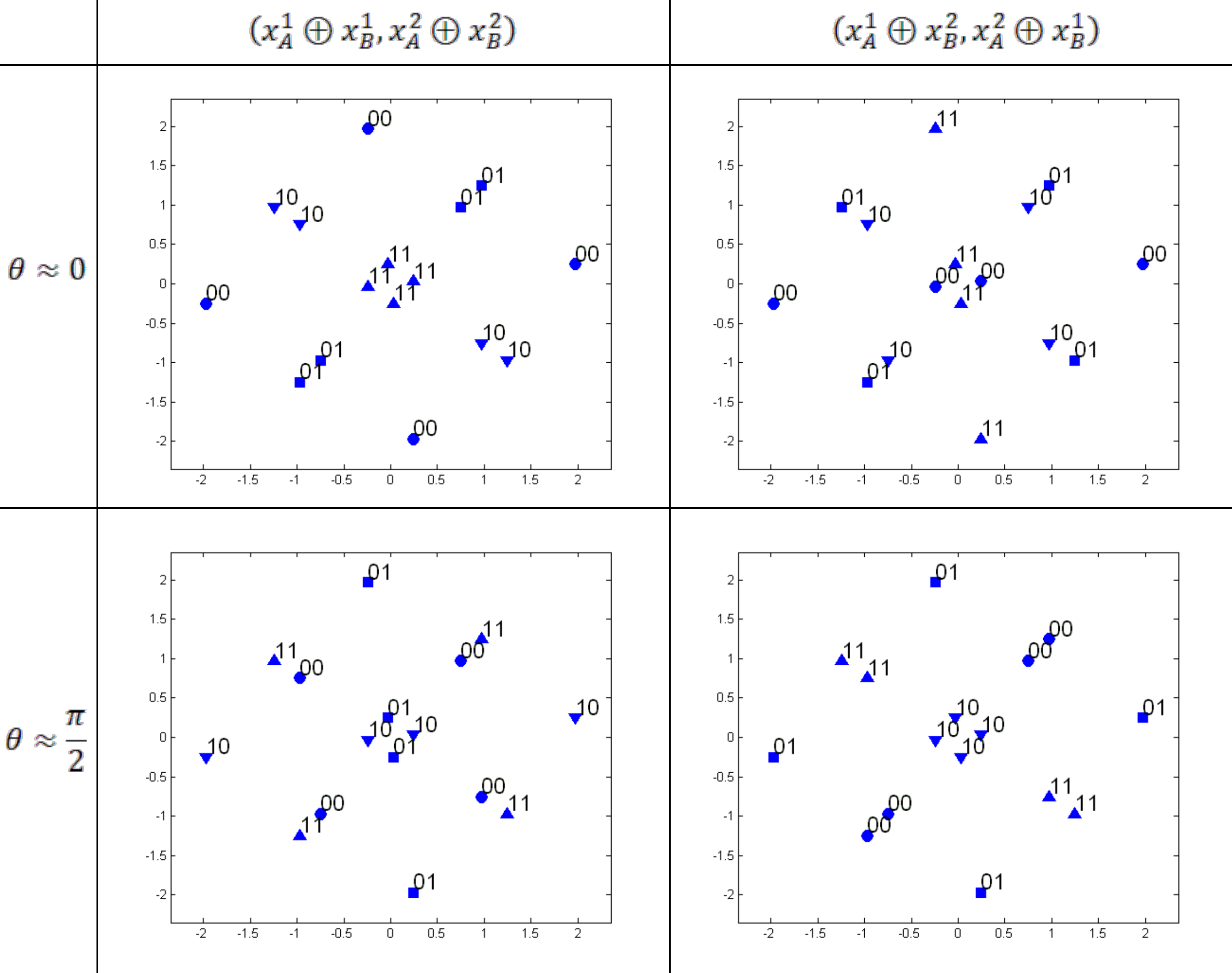}
    \caption{Effective constellation at relay for different values of $\theta$.}

    \label{fig:ConstPlot}
\end{figure*}

For clarification, consider the case of $\ell=2$. Let a function $f_1$ be defined by $\mathbf{D}_A = \mathbf{D}_B =\left[\begin{array}{cc} 1 & 0\\ 0 & 1 \end{array} \right].$ Writing the vectors $\underline{x}_A$ and $\underline{x}_B$ as $[x_A^1 \ x_A^2]^T$ and $[x_B^1 \ x_B^2]^T$ respectively, we see that $\underline{x}_{f_1,R}=[x_{f_1,R}^1 \ x_{f_1,R}^2]^T$ is given by
\begin{equation}
\underline{x}_{f_1,R} = f_1(\underline{x}_A,\underline{x}_B)
=\left[\begin{array}{cccc}
1 & 0 & 1 & 0\\
0 & 1 & 0 & 1
\end{array} \right]
\left[\begin{array}{c}x_A^1 \\ x_A^2 \\ x_B^1 \\ x_B^2 \end{array}\right]. \nonumber
\end{equation}
This corresponds to the binary XOR function given by $f_1(\underline{x}_A,\underline{x}_B)=[x_A^1\oplus x_B^1,x_A^2\oplus x_B^2]^T$.

Define another function $f_2$ using $\mathbf{D}_A = \left[\begin{array}{cc} 1 & 0\\ 0 & 1 \end{array} \right]$ and $\mathbf{D}_B = \left[\begin{array}{cc} 0 & 1\\ 1 & 0 \end{array} \right].$
This is the rotated-XOR function given by $f_2(\underline{x}_A,\underline{x}_B)=[x_A^1\oplus x_B^2,x_A^2\oplus x_B^1]^T$.

Recall from \eqref{eq:xAdef}, that $\underline{v}^k[n] \Leftrightarrow x^k[n]$. Thus using $f_1$ at the relay corresponds to decoding $[\underline{v}_A^1\oplus \underline{v}_B^1]$ and $[\underline{v}_A^2\oplus \underline{v}_B^2]$. Similarly, applying $f_2$ at the relay corresponds to decoding $[\underline{v}_A^1\oplus \underline{v}_B^2]$ and $[\underline{v}_A^2\oplus \underline{v}_B^1]$.

To illustrate the importance of choosing the decoding function $f$ depending on $(h_A,h_B)$, consider an example with $\mathcal{Q}=\{1,j,-1,-j\}=\{\mathcal{M}(00),\mathcal{M}(01),\mathcal{M}(11),\mathcal{M}(10)\}$ (i.e. QPSK with Gray Labeling).  Further, let $h_A = e^{j \theta_A}$ and $h_B = e^{j \theta_B}$, and let $\theta = \theta_A -\theta_B$ be the phase difference. Consider the decoding functions
\begin{align}
f_1(\underline{x}_A,\underline{x}_B)&=[x_A^1\oplus x_B^1,x_A^2\oplus x_B^2] \nonumber \\
f_2(\underline{x}_A,\underline{x}_B)&=[x_A^1\oplus x_B^2,x_A^2\oplus x_B^1]. \nonumber
\end{align}
The resulting constellation $\mathcal{Q}_R$ at the relay is shown for different values of $\theta$ in Fig. \ref{fig:ConstPlot}. Note that the complex coordinates of the constellation points are exactly the same, but their labels are different based on $\theta$ and $f\in\{f_1,f_2\}$. When $\theta\approx 0$, $f_1$ appears to have better performance than $f_2$ in terms of the distances between points with unequal labels. The situation is reversed when $\theta\approx \frac{\pi}{2}$. This shows that the performance for a fixed decoding function can vary widely with $\theta$ even when both $|h_A|$ and $|h_B|$ are large.

As illustrated in Fig. \ref{fig:ConstPlot}, each $f\in\mathcal{F}$ induces a mapping $\mathcal{M}_{f,R}$ between address vectors $\underline{x}_{f,R}$ and constellation points $q_R\in\mathcal{Q}_R$ similar to $\mathcal{M}$ for the point to point case. The relay is only interested in decoding $\mathbf{X}_{f,R}$, which will have $\ell$ rows. Thus $\mathcal{M}_{f,R}$ forms a one-to-many map from length $\ell$ binary address vectors to constellation points. Let $\mathcal{S}\subseteq\{1,...,\ell\}$ be the subset of elements from $\underline{x}_{f,R}$ which are fixed. Then, let $\mathcal{X}_{\{x_{f,R}^k|k\in\mathcal{S}\}}\subseteq\mathbb{F}_2^\ell$ be the subset of $\underline{x}_{f,R}$'s with the same values for all points in $\mathcal{S}$. For a given $f\in\mathcal{F}$, the output of $\mathcal{M}_{f,R}$ is
\begin{align} \label{eq:RelayMapDef}
\mathcal{M}_{f,R}(\{x_{f,R}^k|k\in\mathcal{S}\}) = &\{q_R\in\mathcal{Q}_R|q_R=h_A\mathcal{M}(\underline{x}_A)+h_B\mathcal{M}(\underline{x}_B),~ f(\underline{x}_A,\underline{x}_B)\in\mathcal{X}_{\{x_{f,R}^k|k\in\mathcal{S}\}}\}.
\end{align}
For the example in Fig. \ref{fig:ConstPlot}, $\mathcal{M}_{f,R}(11)$ would return the four constellation points labeled $11$ in each figure. $\mathcal{M}_{f,R}(1?)$ would return the eight constellation points in the union $\mathcal{M}_{f,R}(11)\cup\mathcal{M}_{f,R}(10)$.

In order for nodes A and B to be able to unambiguously decode their desired messages, the authors in \cite{koike2009optimized} show that $f$ must satisfy
\begin{align} \label{eq:unambiguous}
&f(\underline{x}_A,\underline{x}_B) \neq f(\underline{x}_A^\prime ,\underline{x}_B) ~ \forall ~ \underline{x}_A \neq \underline{x}_A^\prime  \textrm{~and~} \underline{x}_B \nonumber \\
&f(\underline{x}_A,\underline{x}_B) \neq f(\underline{x}_A,\underline{x}_B^\prime) ~ \forall ~ \underline{x}_B \neq \underline{x}_B^\prime \textrm{~and~} \underline{x}_A.
\end{align}
We call functions that satisfy this property unambiguous.

\emph{Lemma 1:}
For any $\mathbf{D}_A,\mathbf{D}_B \in \mathbf{\mathcal{D}}$, a decoding function
\begin{equation} \label{eq:RelayFuncDef}
f(\underline{x}_A,\underline{x}_B) = [\mathbf{D}_A \mathbf{D}_B] \left[\begin{array}{c}\underline{x}_A \\ \underline{x}_B \end{array}\right]
\end{equation}
is unambiguous.

\begin{proof}
The proof follows from the invertibility of $\mathbf{D}_A$ and $\mathbf{D}_B$. For some $\underline{x}_A$, suppose that there exists $\underline{x}_B\neq\underline{x}_B^\prime$ so that
\begin{equation}
[\mathbf{D}_A \mathbf{D}_B] \left[\begin{array}{c}\underline{x}_A \\ \underline{x}_B \end{array}\right] = [\mathbf{D}_A \mathbf{D}_B] \left[\begin{array}{c}\underline{x}_A \\ \underline{x}_B^\prime \end{array}\right]. \nonumber
\end{equation}
This can be written as
\begin{align}
\mathbf{D}_A\underline{x}_A\oplus\mathbf{D}_B\underline{x}_B &= \mathbf{D}_A\underline{x}_A\oplus\mathbf{D}_B\underline{x}_B^\prime \nonumber \\
\mathbf{D}_A\underline{x}_A\oplus\mathbf{D}_A\underline{x}_A\oplus\mathbf{D}_B\underline{x}_B &= \mathbf{D}_A\underline{x}_A\oplus\mathbf{D}_A\underline{x}_A\oplus\mathbf{D}_B\underline{x}_B^\prime \nonumber \\
\mathbf{D}_B\underline{x}_B & =\mathbf{D}_B\underline{x}_B^\prime \nonumber \\
\mathbf{D}_B^{-1}\mathbf{D}_B\underline{x}_B & =\mathbf{D}_B^{-1}\mathbf{D}_B\underline{x}_B^\prime \nonumber \\
\underline{x}_B &=\underline{x}_B^\prime \nonumber
\end{align}
which is a contradiction.
\end{proof}

\section{Achievable Information Rates} \label{sec:InfoAnalysis}

\subsection{Achievable Rate for a Given Function}
For a given $f$ and fixed channel gains $h_A$ and $h_B$ the achievable rate region is given by the following theorem. This theorem is the key contribution of this paper.

\emph{Theorem 1:}  Choose some fixed $\mathbf{D}_A,\mathbf{D}_B \in \mathcal{D}$ and define
\begin{equation}
\underline{x}_{f,R} = f(\underline{x}_A,\underline{x}_B) = [\mathbf{D}_A \mathbf{D}_B] \left[\begin{array}{c}\underline{x}_A \\ \underline{x}_B \end{array}\right].
\end{equation}
Choose a subset $\mathcal{S}\subseteq\{1,...,\ell\}$ and define $\overline{\mathcal{S}}=\{1,...,\ell\}\setminus\mathcal{S}$. Divide $\mathcal{S}$ into $p$ non-empty disjoint subsets $\mathcal{S}_1,...,\mathcal{S}_p$ so that $\bigcup_{i=1}^p\mathcal{S}_i = \mathcal{S}$. Let $Z_i,~i\in\{1,...,p\}$ define $p$ i.i.d. Bernoulli random variables with parameter $\frac{1}{2}$. At last, let each row of $\mathbf{X}_A$ and $\mathbf{X}_B$ be encoded using a different coset of the same linear code $\mathcal{C}$. Then there exists a linear code $\mathcal{C}$ of rate $\mathcal{R}$ for which the relay can reliably decode $\mathbf{X}_{f,R}$ as long as $\mathcal{R}$ satisfies
\begin{align} \label{eq:Thm1Bound}
\mathcal{R}< & \underset{\mathcal{S},\overline{\mathcal{S}},\mathcal{S}_1,...,\mathcal{S}_p}{min} ~ \frac{1}{p} I(Y_R;\{X_{f,R}^k|k\in\mathcal{S}\}|\{X_{f,R}^k|k\in\overline{\mathcal{S}}\}, \{X_{f,R}^k\oplus Z_i|k\in\mathcal{S}_i\} ~\forall~ i\in\{1,...,p\}).
\end{align}
For the special case when $\ell=2$, the set of bounds described by \eqref{eq:Thm1Bound} are equivalent to
\begin{align} \label{eq:QPSKRateBound}
\mathcal{R} < min \{&\frac{1}{2}I(Y_R;X_R^1,X_R^2),~I(Y_R;X_R^1|X_R^2),~I(Y_R;X_R^2|X_R^1),~I(Y_R;X_R^1,X_R^2|X_R^1\oplus Z_1,X_R^2\oplus Z_1)\}.
\end{align}

Note that
\[
I(Y_R;X_R^1,X_R^2|X_R^1\oplus Z_1,X_R^2\oplus Z_1)=I(Y_R;X_R^1,X_R^2|X_R^1\oplus X_R^2).
\]
That is, $\{X_R^1\oplus Z_1,X_R^2\oplus Z_1\}$ and $\{X_R^1\oplus X_R^2\}$ carry the same information about $X_R^1$ and $X_R^2$.

\begin{proof}
The detailed proof is provided in the Appendix. However, the key steps in the proof are outlined below.

Our proof uses the standard approach of deriving upper bounds on the probability of error for a joint typicality decoder averaged over a carefully chosen ensemble of codes. The ensemble considered here is the ensemble obtained by using random cosets of the {\em same linear} code for the different signaling levels in the multilevel coding scheme. The linear code is chosen from the ensemble of linear codes with randomly chosen entries in the generator matrix. \emph{The use of the same linear code in each level is an important ingredient in our proposed scheme since we allow the relay to freely take linear combinations of codewords from different signaling levels. However, this is also what complicates the proof.} The ensemble used here is different from the often used ensemble of random coset codes used at each level in the multilevel coding scheme since the latter ensemble allows for independently chosen codes at each level. While the latter ensemble has been used widely to obtain achievable rates for MLC for the point to point channel and the multiple access channel, the former ensemble has not been analyzed in detail in the literature. The key contribution of our proof in the Appendix is to derive the achievable rates with the former ensemble with identical linear codes at each level.

This can be accomplished since the use of the same linear code at each level ensures that for each $f \in \mathcal{F}$, $\mathcal{C}_{f,R}^k,~k\in\{1,...,\ell\}$ is a member of the ensemble used at the transmitters. The main complication that arises from this is that the pairwise independence assertion that is required in typical channel coding proofs \cite{gallager1968information} does not hold for certain classes of error events. Particularly, it is possible for the relay to correctly decode some rows of $\mathbf{X}_{f,R}$ while others may be in error. We handle this by splitting the union bound for error probability into separate classes of error events which are conditionally pairwise independent.

The bound for the $\ell=2$ case can be derived by letting $\mathcal{S},\overline{\mathcal{S}},\mathcal{S}_1,\mathcal{S}_2\subseteq\{1,2\}$ take the following values respectively.
\begin{align}
&\{\mathcal{S}=\{1,2\},\overline{\mathcal{S}}=\emptyset,\mathcal{S}_1=\{1\},\mathcal{S}_2=\{2\}\} \nonumber \\
&\{\mathcal{S}=\{1\},\overline{\mathcal{S}}=\{2\},\mathcal{S}_1=\{1\}\} \nonumber \\
&\{\mathcal{S}=\{2\},\overline{\mathcal{S}}=\{1\},\mathcal{S}_1=\{2\}\} \nonumber \\
&\{\mathcal{S}=\{1,2\},\overline{\mathcal{S}}=\emptyset,\mathcal{S}_1=\{1,2\}\}.
\end{align}
Notice that the first three terms in $\eqref{eq:QPSKRateBound}$ are also required by the proof for multilevel coding for the point to point channel. The last bound is a result of the requirement that each signaling level uses a coset of the same linear code. It would be required for the point to point case as well if the same codes were used at each level.
\end{proof}

It should be noted that the steps of the proof for theorem 1 can applied almost unaltered to the problem of finding the achievable rate for decode-and-forward if nodes A and B transmit using different cosets of the same linear codes at each level. In a decode-and-forward scheme, the relay attempts to reliably decode the messages transmitted from node A and B and then broadcasts a function of the received messages to nodes A and B. With a slight change to the channel model, the proof of theorem 1 can be applied to the problem of recovering the $2\ell$ coset codewords which form the rows of
\begin{equation}
\mathbf{X}_{AB} = \left[\begin{array}{c} \mathbf{X}_A \\ \mathbf{X}_B \end{array}\right].
\end{equation}
Dividing the set $\{1,...,2\ell\}$ into subsets $\mathcal{S},\overline{\mathcal{S}},\mathcal{S}_1,...,\mathcal{S}_p$ as in theorem 1, we can show that $\mathbf{X}_{AB}$ can be reliably decoded as long as $\mathcal{R}$ satisfies
\begin{align} \label{eq:DecForwardBound}
\mathcal{R}< & \underset{\mathcal{S},\overline{\mathcal{S}},\mathcal{S}_1,...,\mathcal{S}_p}{min} ~ \frac{1}{p} I(Y_R;\{X_{AB}^k|k\in\mathcal{S}\}|\{X_{AB}^k|k\in\overline{\mathcal{S}}\}, \{X_{AB}^k\oplus Z_i|k\in\mathcal{S}_i\} ~\forall~ i\in\{1,...,p\}).
\end{align}
Therefore, by allowing the relay to choose between compute-and-forward and decode-and-forward, the maximum of the bounds given by \eqref{eq:Thm1Bound} and \eqref{eq:DecForwardBound} is achievable.

\subsection{Universally Achievable Rate}
We say that a rate $\mathcal{R}$ is universally achievable over the set $\mathcal{H}\subset\mathbb{C}^2$ if there exists a fixed linear code $\mathcal{C}$ of rate $\mathcal{R}$ and coset matrices $\mathbf{\Lambda}_A$ and $\mathbf{\Lambda}_B$ such that for every $(h_A,h_B) \in \mathcal{H}$, the relay can reliably decode $\mathbf{X}_{f,R}$ for some $f\in\mathcal{F}$. That is some $\mathbf{X}_{f,R}$ can be decoded with arbitrarily small probability of error in the usual information-theoretic sense. The main result in this section is the following theorem.

\emph{Theorem 2:} For a fixed $f\in\mathcal{F}$ and $(h_A,h_B)$, define $\mathcal{R}_f(h_A,h_B)$ as the supremum of rates satisfying \eqref{eq:Thm1Bound} where $\underline{x}_{f,R}=f(\underline{x}_A,\underline{x}_B)$. For any finite set of channel gains, $\mathcal{H}\subset\mathbb{C}^2$, any rate $\mathcal{R}$ such that
\begin{equation} \label{eq:Thm2Bound}
\mathcal{R} < \underset{(h_A,h_B)\in\mathcal{H}}{min} ~ \underset{f\in\mathcal{F}}{max} ~ \mathcal{R}_f(h_A,h_B)
\end{equation}
is universally achievable.

\begin{proof}
For a fixed finite $\mathcal{H}\subset\mathbb{C}^2$ and set of decoding functions $\mathcal{F}$, define $\mathcal{R}^{\prime}$ as the supremum of rates satisfying \eqref{eq:Thm2Bound}. Define $\delta>0$ as the acceptable probability of error for a finite length code and choose a fixed $\mathcal{R}<\mathcal{R}^{\prime}$.

We will first consider an arbitrary $(h_A,h_B)\in\mathcal{H}$ and $f\in\mathcal{F}$ such that $\mathcal{R}<\mathcal{R}_f(h_A,h_B)$. Define $\Omega^N$ as the set of coset codes of the form $\{\mathcal{C},\mathbf{\Lambda}_A,\mathbf{\Lambda}_B\}$ which have length $N$. Thus, by increasing the value of $N$ we form a sequence of ensembles of coset codes $\Omega$. Define $P(Err|\Omega^N)$ as the ensemble average probability of decoding error for the ensemble $\Omega^N$. Define $P(Err|\mathcal{C},\mathbf{\Lambda}_A,\mathbf{\Lambda}_B)$ as the probability of decoding error for a specific coset code.

Define $\Omega_{bad}^N\subset\Omega^N$ as
\begin{equation}
\Omega_{bad}^N = \{\mathcal{C},\mathbf{\Lambda}_A,\mathbf{\Lambda}_B\in\Omega^N|P(Err|\mathcal{C},\mathbf{\Lambda}_A,\mathbf{\Lambda}_B)\geq\delta\}.
\end{equation}
Then let $\Omega_{good}^N=\Omega^N\setminus\Omega_{bad}^N$. Define $P(bad|N)=\frac{|\Omega_{bad}^N|}{|\Omega^N|}$ and $P(good|N)=\frac{|\Omega_{good}^N|}{|\Omega^N|}$ as the probability that a bad or good code is selected uniformly at random from $\Omega^N$ respectively. We know that
\begin{align}
P(Err|\Omega^N) &= P(bad|N)P(Err|\Omega_{bad}^N) + P(good|N)P(Err|\Omega_{good}^N) \nonumber \\
&\geq P(bad|N)\delta + P(good|N)P(Err|\Omega_{good}^N) \nonumber \\
&\geq P(bad|N)\delta. \nonumber
\end{align}
The proof of theorem 1 relies on showing that $\underset{N\rightarrow\infty}{lim}P(Err|\Omega^N)=0$. Therefore, there exists some $N_0$ such that for any $N>N_0$,
\begin{align}
&P(bad|N)\delta \leq P(Err|\Omega^N) < \frac{\delta}{\tau} ~\Rightarrow~ P(bad|N) < \frac{1}{\tau} \nonumber
\end{align}
for some finite $\tau>2|\mathcal{H}|$. This means that $|\Omega_{bad}^N|<\frac{|\Omega^N|}{\tau}$. Note that choosing $\tau>2|\mathcal{H}|$ is arbitrary but ensures that $\tau$ will be ``large enough'' to complete the proof.

We want to show the existence of some fixed $\{\mathcal{C},\mathbf{\Lambda}_A,\mathbf{\Lambda}_B\}\in\Omega^N$ such that for every $(h_A,h_B)\in\mathcal{H}$ there is some $f\in\mathcal{F}$ so that $P(Err|\mathcal{C},\mathbf{\Lambda}_A,\mathbf{\Lambda}_B)<\delta$. We can apply the steps above to find a set $\Omega_{bad}^N(h_A,h_B)$ for every $(h_A,h_B)\in\mathcal{H}$. Since $|\mathcal{H}|$ is finite the largest $N$ required by any $(h_A,h_B)\in\mathcal{H}$ must exist and be a finite integer $N_{max}$.

Since $\tau$ is chosen to be larger than $2|\mathcal{H}|$, the set
\[
\Omega^{N_{max}}\setminus\{\bigcup_{(h_A,h_B)\in\mathcal{H}}\Omega_{bad}^{N_{max}}(h_A,h_B)\}
\]
must be non-empty because
\[
\sum_{(h_A,h_B)\in\mathcal{H}}|\Omega_{bad}^{N_{max}}(h_A,h_B)|\leq|\Omega^{N_{max}}|/2.
\]
Thus, since at least half of the codes are always good, there exists at least one coset code which allows reliable decoding for every $(h_A,h_B)\in\mathcal{H}$ as long as $\mathcal{R}<\mathcal{R}^{\prime}$.

\end{proof}

%

Note that in order for this problem to be practically interesting, the set $\mathcal{H}$ should be meaningfully defined. It may seem more natural to evaluate our scheme based on the outage probability for a fixed transmission rate. We consider the universally achievable rate formulation for two reasons. First, the outage probability can be determined for the block fading channel using results from Theorems 1 and 2. Second, the universally achievable rate is useful for illustrating the flexibility of the proposed scheme to phase mismatch between nodes A and B. This is especially interesting if we consider a system where the relay is used to provide power control information to nodes A and B as in \cite{wilson2009power}.

\section{Numerical Results} \label{sec:NumericalResults}

\subsection{Numerical Results for QPSK}
\begin{figure}
    \centering
    \includegraphics[width=3.5in]{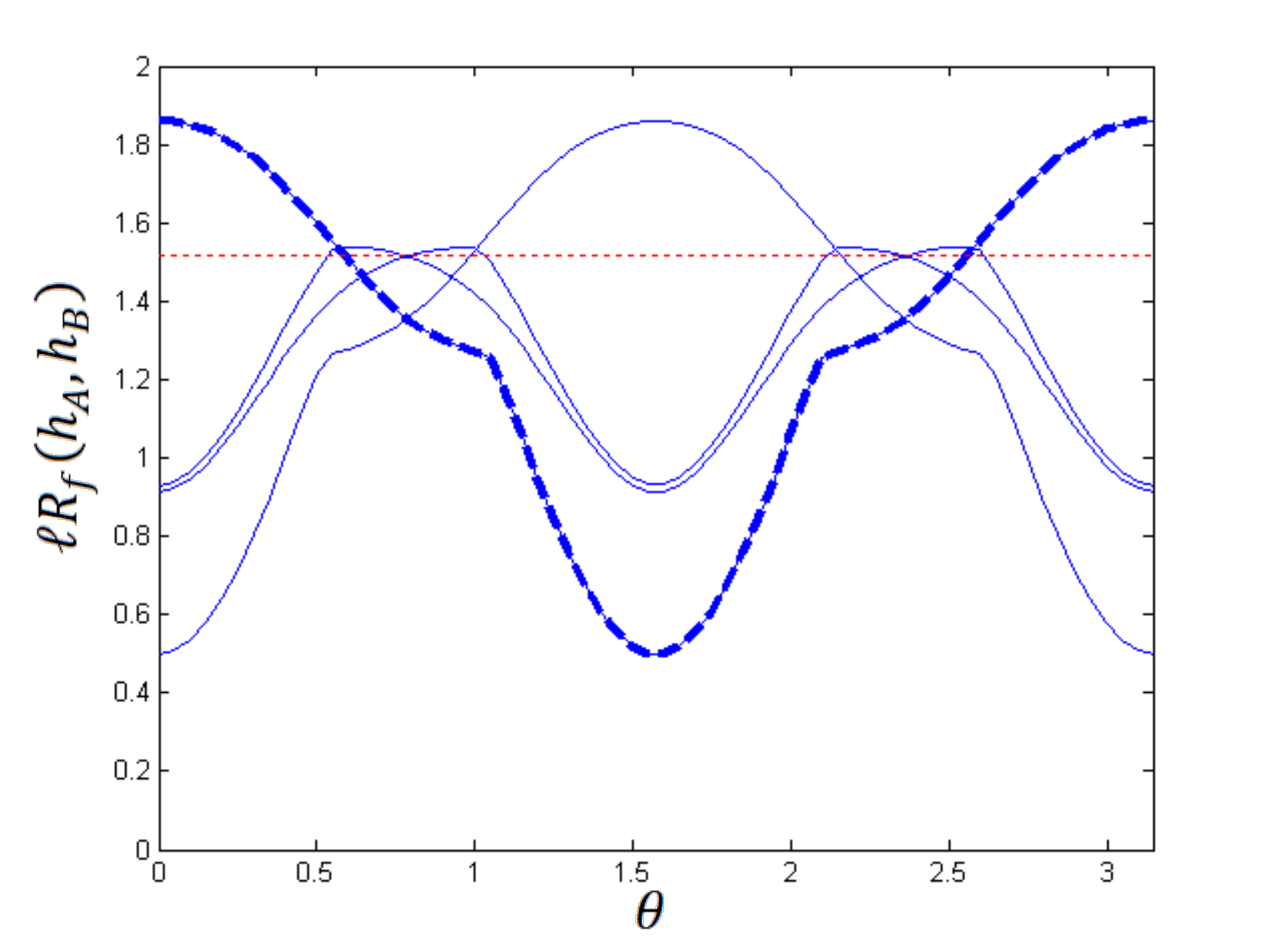}
    \caption{$\ell\mathcal{R}_f(h_A,h_B)$ vs. $\theta$ for each $f\in\mathcal{F}$.}

    \label{fig:Iyx_vs_theta_MLC}
\end{figure}

\begin{figure}
    \centering
    \includegraphics[width=3.5in]{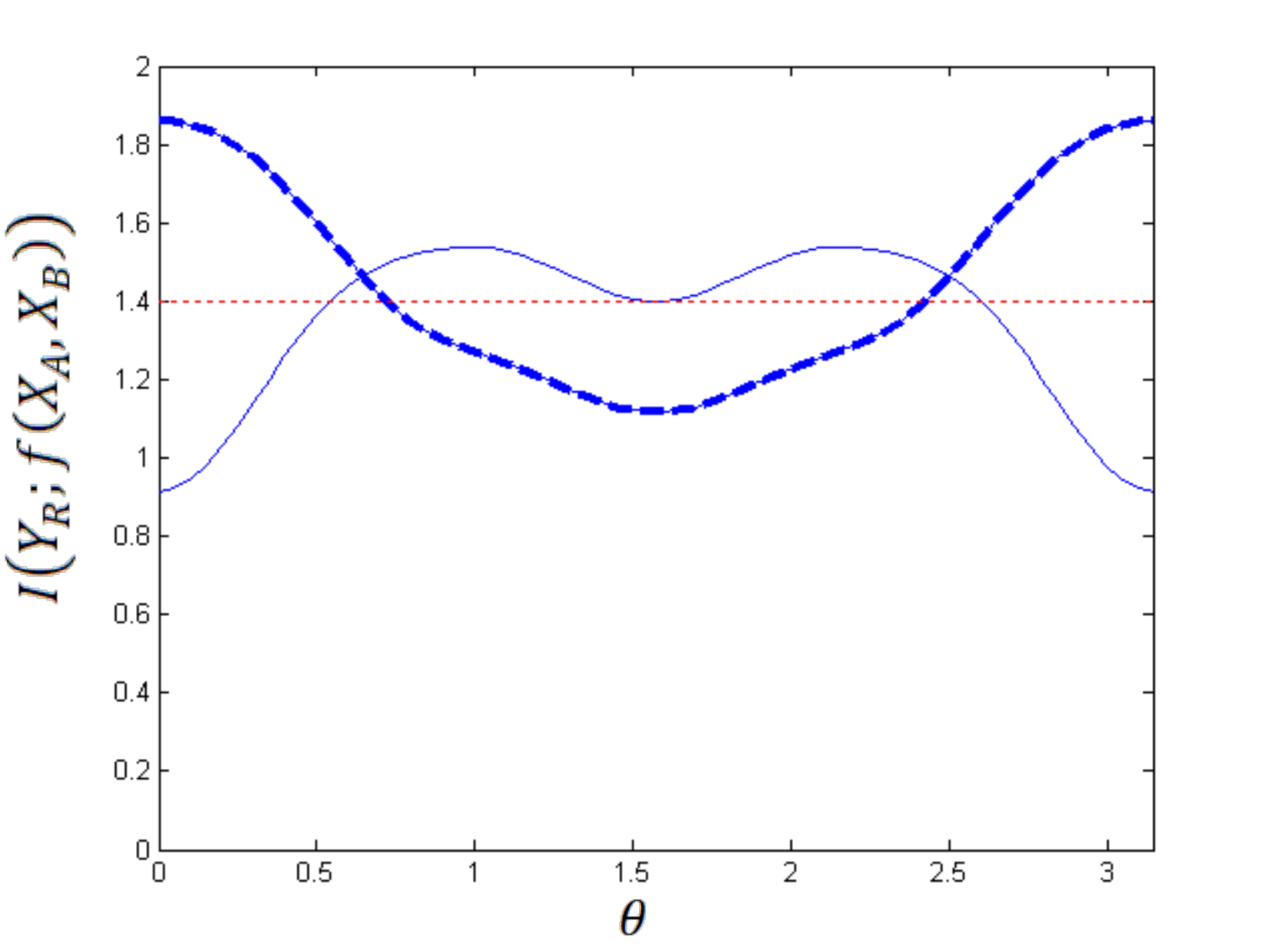}
    \caption{$I(Y_R;f(X_A,X_B))$ vs. $\theta$ for each $f\in\mathcal{F}_{GF4}$.}

    \label{fig:Iyx_vs_theta_GF4}
\end{figure}
As an example, consider the case where nodes A and B transmit symbols from a QPSK constellation with Gray Labeling. Fig.~\ref{fig:Iyx_vs_theta_MLC} shows a plot of the achievable information rate $\ell \mathcal{R}_f(h_A,h_B)$ as given in \eqref{eq:QPSKRateBound} for each function $f \in \mathcal{F}$ dependent on the phase difference $\theta = \theta_A-\theta_B$ for an SNR of $7~dB$. $\mathcal{H}$ is the set of channel gains
\begin{equation} \label{eq:HSetDef}
\mathcal{H} = \{(h_A,h_B)|h_A=e^{j\theta_A},h_B=e^{j\theta_B}\}
\end{equation}
where $\theta_A,\theta_B\in\{0,\frac{\pi}{m},...,2\pi\}$ for a finite integer $m$. Thus $|\mathcal{H}|$ is finite but approximates the selection of any value of $\theta_A$ and $\theta_B$ arbitrarily closely.

The dotted line indicates the universally achievable rate in bits per complex symbol for the proposed scheme which satisfies Theorem 2 for $\mathcal{H}$. Note that different functions provide the best performance for different values of $\theta$ which reiterates the substantial benefit of decoding adaptively.
Notice that a small increase in rate makes reliable decoding impossible for any $f\in\mathcal{F}$ for a significant range of $\theta$; however, there are many $(h_A,h_B)\not\in\mathcal{H}$ such that $\exists f\in\mathcal{F}$ for which reliable decoding is possible.

\subsection{Coding over $GF(4)$}
It is interesting to use this QPSK example to compare our MLC scheme to the case where nodes A and B encode their a data using a linear code $\mathcal{C}_{GF4}$ over $\mathbb{F}_4$ of rate $\mathcal{R}_{GF4}$. The relay uses the set of decoding functions $\mathcal{F}_{GF4}$ corresponding to linear combinations of codewords in $\mathbb{F}_4$ of the form
\begin{equation}
\underline{v}_R = f(\underline{v}_A,\underline{v}_B) = \alpha \underline{v}_A \oplus \beta \underline{v}_B ,~ \alpha,\beta\in \mathbb{F}_4\backslash\{0\}.
\end{equation}
Node A can decode $\underline{v}_B$ from $\underline{v}_A$ and $\underline{v}_R$ by
\begin{equation}
\underline{v}_B = \beta^{-1}(\alpha \underline{v}_A)\oplus \underline{v}_R.
\end{equation}
Node B can recover $\underline{v}_A$ similarly. The relay should be able to decode $\underline{v}_R$ reliably as long as there exists some $f\in\mathcal{F}_{GF4}$ for which
\begin{equation}
\mathcal{R}_{GF4} < I(Y_R;f(X_A,X_B)).
\end{equation}
The value of $I(Y_R;f(X_A,X_B))$ for each possible $f\in\mathcal{F}_{GF4}$ is plotted as a function of $\theta$ in Fig. \ref{fig:Iyx_vs_theta_GF4} with an SNR of $7~dB$. Again the dotted line represents universally achievable rate for the $\mathcal{H}$ in \eqref{eq:HSetDef}.

\subsection{Comparison of Proposed Techniques}
These numerical results illustrate that the proposed MLC scheme facilitates better decoding flexibility at the relay than coding over $\mathbb{F}_4$ for this example. In fact, in an analysis of these functions based on the labeling of points in $\mathcal{Q}_{R}$, it can be seen that $\mathcal{F}_{GF4}\subset\mathcal{F}$. However, this improved flexibility comes at the cost of additional rate constraints on each $f\in\mathcal{F}$. The thick dashed line in Figs. \ref{fig:Iyx_vs_theta_MLC} and \ref{fig:Iyx_vs_theta_GF4} represents the rate which is achievable if the relay decodes using some $f$ which is equivalent to the componentwise xor operation for multilevel coding or finite field addition for $\mathbb{F}_4$. The difference between these curves illustrates the effects of the additional rate constraints imposed by \eqref{eq:Thm1Bound}. In Fig. \ref{fig:Iyx_vs_theta_MLC} the last term $I(Y_R;X_R^1,X_R^2|X_R^1\oplus Z_1,X_R^2\oplus Z_1)$ in \eqref{eq:QPSKRateBound} is dominant if $\theta\approx\frac{\pi}{2}$ for determining the achievable rate for this function. In Fig. \ref{fig:Iyx_vs_theta_GF4} we see that this term does not need to be satisfied if nodes A and B use a linear code in $\mathbb{F}_4$.

\begin{figure}
    \centering
    \includegraphics[width=3.5in]{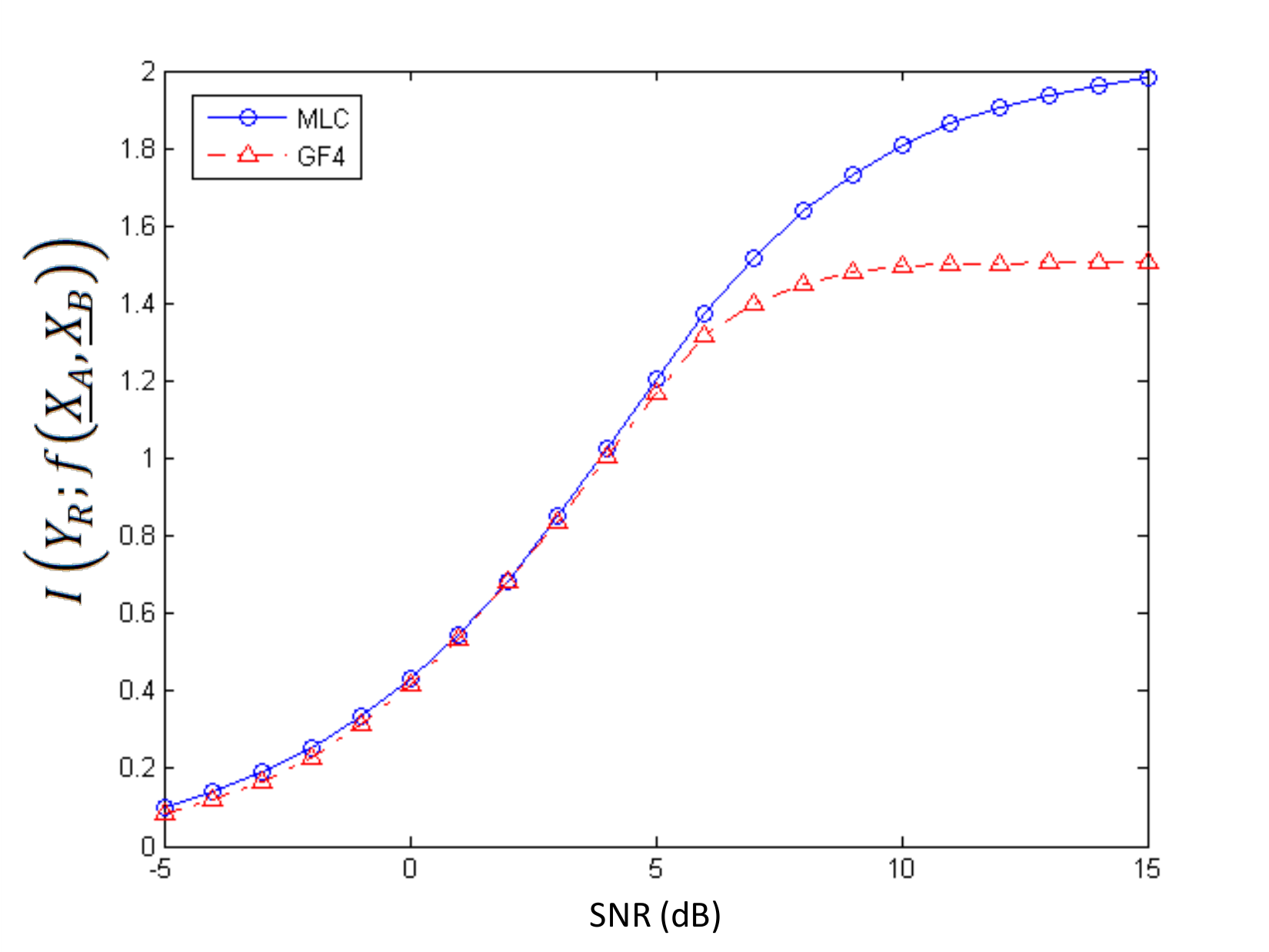}
    \caption{Universally achievable rates vs. SNR(dB) for proposed relaying techniques}

    \label{fig:Iyx_vs_snr}
\end{figure}

The universally achievable rate for the $\mathcal{H}$ in \eqref{eq:HSetDef} (i.e. the constant value given by the dotted line in Figs. \ref{fig:Iyx_vs_theta_MLC} and \ref{fig:Iyx_vs_theta_GF4}) is plotted as a function of SNR in Fig. \ref{fig:Iyx_vs_snr} for the cases where the relay uses $\mathcal{F}$ or $\mathcal{F}_{GF4}$. This value asymptotically approaches 1.5 bits per symbol for coding over $\mathbb{F}_4$. From Fig. \ref{fig:Iyx_vs_theta_GF4}, this appears to occur because $\mathcal{F}_{GF4}$ does not provide the relay with a decoding function which works well when $\theta\approx\frac{\pi}{2}$. This represents an extreme case, because the event $|h_A|=|h_B|$ occurs with probability zero for many random fading processes. However, this illustrates that for PLNC it is possible for the universally achievable rate to be limited by specific $(h_A,h_B)\in\mathcal{H}$ even if each $|h_A|$ and $|h_B|$ is large.

\begin{figure}
    \centering
    \includegraphics[width=3.5in]{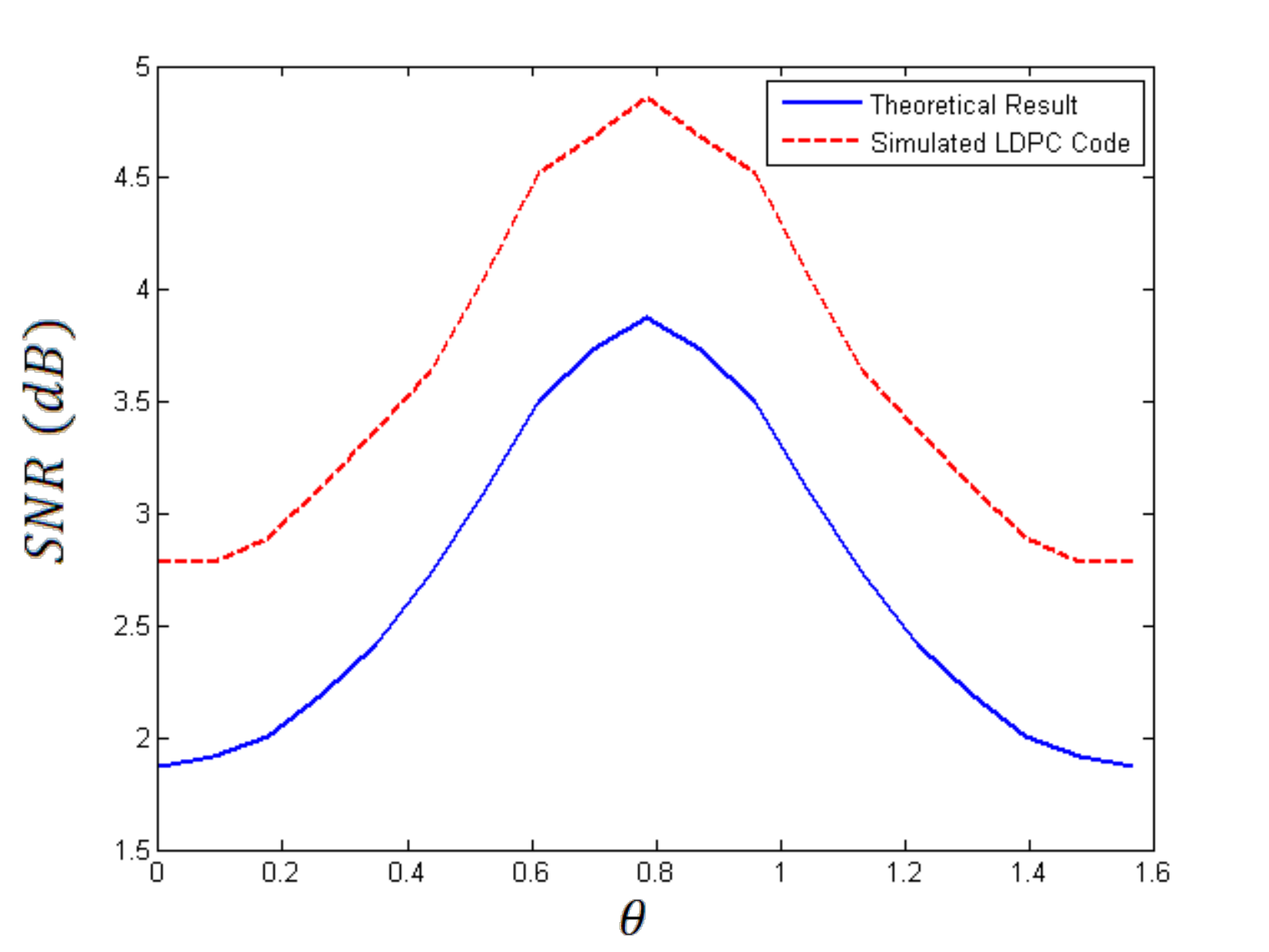}
    \caption{Required SNR(dB) vs. $\theta$ to reliably decode a rate $\frac{1}{2}$ code}
    \label{fig:LDPC3_6_code}
\end{figure}

\subsection{Simulation Results}
To corroborate these theoretical results, we simulated the performance of a regular (3,6) low density parity check (LDPC) code. In Fig. \ref{fig:LDPC3_6_code} the required SNR for a rate $\frac{1}{2}$ code is plotted as a function of $\theta=\theta_A-\theta_B$ for the case where $(h_A,h_B)\in\mathcal{H}$ for the $\mathcal{H}$ in \eqref{eq:HSetDef}. The solid curve represents the theoretically required SNR as determined by \eqref{eq:Thm1Bound}. The dashed red curve represents the SNR for which zero bit errors occurred during 200 simulations of a length $10^5$ code for each tested $\theta$. For a point to point Gaussian channel using binary phase shift keying, it has been shown in \cite{richardson2008modern} that the required SNR for a (3,6) LDPC code with iterative decoding is about 1dB away from the Shannon limit for the same channel and modulation format. In Fig. \ref{fig:LDPC3_6_code}, we see that this trend appears to hold for our scheme as well. We leave more rigorous testing for future work. Note that to achieve the theoretical limit imposed by \eqref{eq:Thm1Bound} using structured codes, it will be necessary to design coding schemes which universally achieve the capacity for many channel conditions. It appears that the class of spatially coupled LDPC codes would be a good choice for this \cite{yedla2011universality}.

\section{Concluding Remarks} \label{sec:ConcludingRemarks}
In this paper, we have proposed a coding scheme based on MLC for compute and forward or PLNC for the case when the channel is perfectly estimated at each receiver but unknown to each transmitter. We showed that MLC allows for decoding of a set of functions of the transmitted messages and the relay can choose one function from this set depending on the channel coefficients. In Theorem~1, we obtained an achievable rate for a fixed decoding function and channel realizations. In Theorem~2, we obtained a numerically computable expression for the universally achievable information rate over a set of channel realizations. Numerical results for QPSK suggest that the proposed scheme significantly outperforms the use of a fixed decoding function with binary linear codes and is better than using linear codes over $\mathbb{F}_4$.

\appendix[Proof of Theorem 1]
Theorem 1 states that for a fixed $f\in\mathcal{F}$, if each $\mathcal{C}_A^1,...,\mathcal{C}_A^\ell$ and $\mathcal{C}_B^1,...,\mathcal{C}_B^\ell$ is a different coset of the same linear code $\mathcal{C}$ of rate $\mathcal{R}$, then there exists some $\mathcal{C}$ for which the relay can reliably decode $\mathbf{X}_{f,R}$ for a suitably chosen $\mathcal{R}$.


\subsection{Additional Notation}
A few definitions only necessary for this proof have been omitted from the main text but are included here for clarity.

We refer to the noiseless observed sequence at the relay as
\begin{equation} \label{eq:qRDef}
\underline{q}_R=h_A\underline{s}_A+h_B\underline{s}_B.
\end{equation}
Where $\underline{q}_R\in\mathcal{Q}_R^N$. Thus the relay observes the noisy observations
\begin{equation} \label{eq:yRDef}
\underline{y}_R = h_A\underline{s}_A + h_B\underline{s}_B+\underline{w}_R = \underline{q}_R + \underline{w}_R.
\end{equation}

When it is necessary to refer to variables associated with different messages, we will refer to variables like $\mathbf{U}_{j,A}$ by the integer $j\in\{0,...,2^{K\ell}-1\}$, whose binary expansion is given by node A's unparsed message $\underline{u}_A$. We assume that nodes A and B encode $\mathbf{U}_{j_A,A}$ and $\mathbf{U}_{j_B,B}$ for transmission, and that the relay observes the noisy samples corresponding to $\mathbf{X}_{j_R,f,R}$. Note that the index $j_R$ of the desired message is a function of $j_A$, $j_B$, and $f$.

The relay will attempt to reliably decode $\mathbf{X}_{j_R,f,R}$ from $\underline{y}_R$ using a joint typicality decoder. Thus the decoder declares an error if either $\mathbf{X}_{j_R,f,R}$ is not jointly typical with $\underline{y}_R$ or if some incorrect message $\mathbf{X}_{j,f,R},~j\neq j_R$ is jointly typical with $\underline{y}_R$. We derive an upper bound on the error probability for this decoder over the ensemble of random coset codes. Specifically, let the elements of $\mathbf{G}$, $\mathbf{\Lambda}_A$, and $\mathbf{\Lambda}_B$ be i.i.d. Bernoulli random variables with parameter $\frac{1}{2}$.

\subsection{Pairwise Independence of Codewords}
Here we provide a brief analysis of the ensemble of coset codes used by nodes A and B and observed by the relay. The following lemmas are stated as they pertain to a nameless encoder to simplify notation. Both lemmas appear as part of the proof of Gallager's Coding Theorem for Random Parity Check Codes \cite{gallager1968information}. We include these proofs because the intuition behind some of the steps is used for other parts of the proof of Theorem 1.

\emph{Lemma 2:} Let each element of $\mathbf{G}$ and $\underline{\lambda}^k$ be i.i.d. Bernoulli random variables with parameter $\frac{1}{2}$. Then we have
\begin{equation}
P(\underline{V}_{j}^k=\underline{v}_{j}^k) = \frac{1}{2^{N}} ~\forall~ \underline{v}_{j}^k\in\mathbb{F}_2^N.
\end{equation}
That is, the codeword $\underline{v}_{j}^k\in\mathbb{F}_2^N$ associated with message vector $\underline{u}_{j}^k\in\mathbb{F}_2^K$ can take any value with uniform probability over the ensemble of random coset codes.

\begin{proof}
For a fixed $\mathbf{G}$ and $\underline{u}_{j}^k$, the output of the linear encoder $\underline{\gamma}_{j}^k=\underline{u}_{j}^k\mathbf{G}$ must take some value in $\mathbb{F}_2^N$. Since $\underline{\lambda}^k$ can take any value with equal probability we have
\begin{align}
P(\underline{V}_{j}^k = \underline{v}_{j}^k) = P(\underline{\Lambda}^k=\underline{v}_{j}^k\oplus\underline{\gamma}_{j}^k) =\frac{1}{2^{N}}~\forall~\underline{v}_{j}^k\in\mathbb{F}_2^N.
\end{align}
\end{proof}

\emph{Lemma 3:} Let each element of $\mathbf{G}$ and $\underline{\lambda}^k$ be i.i.d. Bernoulli random variables with parameter $\frac{1}{2}$. Then for any $j^\prime\neq j$ for which $\underline{u}_{j}^k\neq\underline{u}_{j^\prime}^k$, we have
\begin{align}
P(\underline{V}_{j}^k=\underline{v}_{j}^k,\underline{V}_{j^\prime}^k=\underline{v}_{j^\prime}^k)
&= P(\underline{V}_{j}^k=\underline{v}_{j}^k)P(\underline{V}_{j^\prime}^k=\underline{v}_{j^\prime}^k) \nonumber \\
&= \frac{1}{2^{2N}} ~\forall~ \underline{v}_{j}^k,\underline{v}_{j^\prime}^k\in\mathbb{F}_2^N.
\end{align}
That is, the codewords $\underline{v}_{j}^k~\textrm{and}~\underline{v}_{j^\prime}^k$ associated with $\underline{u}_{j}^k~\textrm{and}~\underline{u}_{j^\prime}^k$ respectively are pairwise independent and uniformly distributed over $\mathbb{F}_2^N$.

\begin{proof}
Suppose that $\underline{u}_{j}^k$ and $\underline{u}_{j^\prime}^k$ differ in the $m_{th}$ position, and let $\underline{g}_i,~i\in\{1,...,K\}$ refer to the $i_{th}$ row of $\mathbf{G}$. Then for any set of rows
\[\underline{g}_1,...,\underline{g}_{m-1},\underline{g}_{m+1},...,\underline{g}_K \]
there is some $\underline{g}_m$ which gives $\underline{v}_{j}^k\oplus\underline{v}_{j^\prime}^k=\underline{\gamma}_{j}^k\oplus\underline{\gamma}_{j^\prime}^k$ any fixed value. By the construction of $\mathbf{G}$ and Lemma 2, $\underline{g}_m$ and $\underline{v}_{j}^k$ can take any value with uniform probability. We can conclude that
\begin{align} \label{eq:PairIndep}
&P(\underline{V}_{j}^k=\underline{v}_{j}^k,\underline{V}_{j^\prime}^k=\underline{v}_{j^\prime}^k |\underline{u}_{j}^k\neq\underline{u}_{j^\prime}^k) \nonumber \\
&= P(\underline{V}_{j}^k=\underline{v}_{j}^k|\underline{u}_{j}^k\neq\underline{u}_{j^\prime}^k)
P(\underline{V}_{j^\prime}^k=\underline{v}_{j^\prime}^k|\underline{v}_{j}^k,\underline{u}_{j}^k\neq\underline{u}_{j^\prime}^k) \nonumber \\
&= P(\underline{\Lambda}^k=\underline{v}_{j}^k\oplus\underline{\gamma}_{j}^k|\underline{u}_{j}^k\neq\underline{u}_{j^\prime}^k)
P(\underline{G}_m=\underline{v}_{j^\prime}^k\oplus\underline{v}_{j}^k|\underline{v}_{j}^k,\underline{u}_{j}^k\neq\underline{u}_{j^\prime}^k) \nonumber \\
&= \frac{1}{2^N}\frac{1}{2^N} = \frac{1}{2^{2N}} ~\forall~ \underline{v}_{j}^k,\underline{v}_{j^\prime}^k\in\mathbb{F}_2^N.
\end{align}
\end{proof}

The key idea behind each proof is the same. In Lemma 2, we see that the uniform distribution of $\underline{\lambda}^k$ implies the uniform distribution of $\underline{v}_{j}^k$. In Lemma 3, we see that the uniform distribution of $\mathbf{G}$ implies the pairwise independence of codewords corresponding to distinct messages.

\subsection{Distribution of Received Signal}
The last step before deriving upper bounds on the probability of decoding error is to derive the distribution of the signal received by the relay over the ensemble of codes.

\emph{Lemma 4:} Let $\mathbf{T}\in\mathbb{F}_2^{m\times2\ell}$ be a matrix of rank $m$, and let $\underline{x}_{\mathbf{T}}[n]\in\mathbb{F}_2^m$ be defined by
\begin{equation}
\underline{x}_{\mathbf{T}}[n] = \mathbf{T} \left[\begin{array}{c} \underline{x}_{j_A,A}[n] \\ \underline{x}_{j_B,B}[n] \end{array}\right].
\end{equation}
For a fixed $\mathbf{T}$, define the set $\mathcal{Q}_{\underline{x}_{\mathbf{T}},R}$ by
\begin{align}
\mathcal{Q}_{\underline{x}_{\mathbf{T}},R} = \{q_R\in\mathcal{Q}_R&|q_R = h_A\mathcal{M}(\underline{x}_A)+h_B\mathcal{M}(\underline{x}_B) ~\textrm{for some}~ \underline{x}_A,\underline{x}_B \in \mathbb{F}_2^\ell ~\textrm{such that}~ \underline{x}_{\mathbf{T}}[n] = \mathbf{T} \left[\begin{array}{c} \underline{x}_A \\ \underline{x}_B \end{array}\right]\}.
\end{align}
Then the distribution of $\underline{q}_R[n]$ conditioned on $\underline{x}_{\mathbf{T}}[n]$ is given by
\begin{equation} \label{eq:qRCondDist}
P(\underline{Q}_{R}[n]=q_R|\underline{x}_{\mathbf{T}}[n]) = \frac{1}{2^{2\ell-m}} ~\forall~ q_R\in\mathcal{Q}_{\underline{x}_{\mathbf{T}}[n],R}.
\end{equation}

\begin{proof}
For a given $\mathbf{U}_{j_A,A}$ and $\mathbf{U}_{j_B,B}$, each $\underline{v}_{j_A,A}^k,\underline{v}_{j_B,B}^k,~k\in\{1,...,\ell\}$ is uniformly distributed in $\mathbb{F}_2^N$ by Lemma 2. Thus over the ensemble of codes, each element of $\mathbf{X}_{j_A,A}$ and $\mathbf{X}_{j_B,B}$ is an i.i.d. Bernoulli random variable with parameter $\frac{1}{2}$.  By \eqref{eq:qRDef}, we have
\begin{equation}
\underline{q}_R[n] = h_A\mathcal{M}(\underline{x}_{j_A,A}[n])+h_B\mathcal{M}(\underline{x}_{j_B,B}[n]).
\end{equation}
Thus for fixed channel gains, $\underline{q}_R[n]$ is a bijective function of $\underline{x}_{j_A,A}[n]$ and $\underline{x}_{j_B,B}[n]$. For a given $\underline{x}_{\mathbf{T}}[n]$ we have,
\begin{align}
P(\underline{X}_{j_A,A}[n]&=\underline{x}_A,\underline{X}_{j_B,B}[n]=\underline{x}_B|\underline{x}_{\mathbf{T}}[n]) = \frac{1}{2^{2\ell-m}} ~\forall~ \underline{x}_A,\underline{x}_B \in \mathbb{F}_2^\ell ~\textrm{such that}~ \underline{x}_{\mathbf{T}}[n]=\mathbf{T} \left[\begin{array}{c} \underline{x}_A \\ \underline{x}_B \end{array}\right]. \nonumber
\end{align}
That is the distribution of $2\ell$ i.i.d. Bernoulli random variables with parameter $\frac{1}{2}$ conditioned on $m$ linear combinations of these variables is uniform over the space of outcomes satisfying the linear constraints. The result follows because $\underline{q}_R[n]$ is a function of the address vectors.
\end{proof}

By \eqref{eq:yRDef}, the conditional distribution of $\underline{y}_R[n]$ on $\{x_{j_R,f,R}^k[n]|k\in\mathcal{S}\}$ is given by
\begin{align} \label{eq:yRCondDist}
&P(y_R[n]|\{x_{j_R,f,R}^k[n]|k\in\mathcal{S}\}) = \frac{1}{2^{2\ell-|\mathcal{S}|}}\sum_{q_R\in\mathcal{M}_{f,R}(\{x_{j_R,f,R}^k[n]|k\in\mathcal{S}\})}\frac{1}{\sqrt{2\pi\sigma^2}} e^{-\frac{|y_R[n]-q_R|^2}{2\sigma^2}}
\end{align}
where $\mathcal{M}_{f,R}$ is the mapping function at the relay induced by $f$ and the channel conditions.

\subsection{Analysis of Error Probability}
The relay uses a joint typicality decoder to decode $\mathbf{X}_{j_R,f,R}$ from $\underline{y}_R$. For some fixed $\epsilon>0$, define $\mathcal{A}_\epsilon^N$ as the set of $(\mathbf{X}_{j,f,R},\underline{y}_R)$ pairs which satisfy the definition of joint typicality given in \cite{cover2006elements}. The set $\mathcal{A}_\epsilon^N$ is referred to as the jointly typical set. Let the event $E_{j},~j\in\{0,...,2^{K\ell}-1\}$ be the event $(\mathbf{X}_{j,f,R},\underline{y}_R)\in \mathcal{A}_\epsilon^N$. The probability of error given that the codeword corresponding to $j_R$ is observed by the relay can be expressed
\begin{equation}
P(Err|j_R) = P(\overline{E}_{j_R}\cup \bigcup_{j\neq j_R}^{2^{K\ell}-1}E_j|j_R).
\end{equation}
Applying the union bound, we get
\begin{equation} \label{eq:PerrUnionBound}
P(Err|j_R) \leq P(\overline{E}_{j_R}|j_R) + \sum_{j\neq j_R}^{2^{K\ell}-1} P(E_j|j_R).
\end{equation}

Recall that $\underline{y}_R$ is the result of the relay observing the symbol sequence associated with message $j_R$. Thus by the joint asymptotic equipartition property (AEP) we have that for any $\epsilon>0$,
\begin{equation}
P(\overline{E}_{j_R}|j_R) < \epsilon
\end{equation}
for sufficiently large $N$.

The proof of the channel coding theorem for the general discrete memoryless channel in \cite{cover2006elements} relies on upper bounding $P(E_j|j_R)$ using the joint AEP. This is not straightforward here because $\mathbf{X}_{j,f,R},~j\neq j_R$ and $\underline{y}_R$ are not independent with the same marginals for certain classes of error events. For example, if $\ell=2$, we could have
\begin{align}
\mathbf{U}_{j,f,R} &= \mathbf{U}_{j_R,f,R}\oplus \left[\begin{array}{c} \underline{e}_{u} \\ \underline{0} \end{array}\right] \nonumber \\
\Rightarrow \mathbf{X}_{j,f,R} &= \mathbf{X}_{j_R,f,R}\oplus \left[\begin{array}{c} \underline{e}_{v} \\ \underline{0} \end{array}\right]
\end{align}
for some $\underline{e}_{u}\in\mathbb{F}_2^{K}\setminus\{0\}$ and $\underline{e}_{v}\in\mathbb{F}_2^{N}\setminus\{0\}$. This means that $\underline{v}_{j,f,R}^1\neq\underline{v}_{j_R,f,R}^1$ but $\underline{v}_{j,f,R}^2=\underline{v}_{j_R,f,R}^2$. Thus for this class of error events $\mathbf{X}_{j,f,R}$ and $\mathbf{X}_{j_R,f,R}$ are not pairwise independent. Note that this class of error events is handled by the proof of the coding theorem for the multiple access channel \cite{wachsmann1999multilevel}, \cite{cover2006elements}, and \cite{gallager1985perspective}. In the coding theorem proof for the multiple access channel, it is possible for the receiver to correctly decode a codeword from one transmitter while making an error in decoding the codeword from a second transmitter. This has the same effect as correctly decoding the codeword on one level of a multilevel encoder while making an error in decoding the codeword transmitted on the second level.

Unfortunately, choosing to use a coset of the \emph{same} linear codes at each bit level introduces a new class of error events of the form
\begin{align}
\mathbf{U}_{j,f,R} &= \mathbf{U}_{j_R,f,R}\oplus \left[\begin{array}{c} \underline{e}_{u} \\ \underline{e}_{u} \end{array}\right] \nonumber \\
\Rightarrow \mathbf{X}_{j,f,R} &= \mathbf{X}_{j_R,f,R}\oplus \left[\begin{array}{c} \underline{e}_{v} \\ \underline{e}_{v} \end{array}\right].
\end{align}
For this class of error events, the columns of the error matrix $\mathbf{X}_{j,f,R}\oplus\mathbf{X}_{j_R,f,R}$ must be in $\{[0~0]^T,[1~1]^T\}$. \emph{This is the key difference between our proof and the proofs for the general multiple access channel or point to point channel with multilevel coding}.

We can move forward by splitting the sum in \eqref{eq:PerrUnionBound} into different events for which $\mathbf{X}_{j,f,R}$ and $\mathbf{X}_{j_R,f,R}$ are conditionally pairwise independent. Define a set of $p\leq \ell$ disjoint subsets $\mathcal{S}_1,...,\mathcal{S}_p\subseteq\{1,...,\ell\}$. Let $t_m$ be the smallest element of $\mathcal{S}_m$, and define the sets $\mathcal{T}=\{t_1,...,t_p\}$, $\mathcal{S}=\mathcal{S}_1\cup...\cup\mathcal{S}_p$, and $\overline{\mathcal{S}}=\{1,...,\ell\}\setminus \mathcal{S}$. For each set of subsets, define an index set $\mathcal{J}_{\mathcal{S}_1,...,\mathcal{S}_p}$ given by
\begin{align} \label{eq:PairwiseSubsets}
\mathcal{J}_{\mathcal{S}_1,...,\mathcal{S}_p} = \left\{j\left|\underline{u}_{j,f,R}^k = \left\{ \begin{array}{ll}
         \underline{u}_{j_R,f,R}^k\oplus\underline{e}_{u,1} & \mbox{, $k \in \mathcal{S}_1$}\\
         &\vdots \\
         \underline{u}_{j_R,f,R}^k\oplus\underline{e}_{u,p} & \mbox{, $k \in \mathcal{S}_p$}\\
         \underline{u}_{j_R,f,R}^k & \mbox{, $k\not\in\mathcal{S}$}\end{array}\right.\right.\right\}.
\end{align}
Here each message error vector, $\underline{e}_{u,i}\in\mathbb{F}_2^K\backslash\{\underline{0}\}~\forall~i\in\{1,...,p\}$ satisfies $\underline{e}_{u,i}\neq\underline{e}_{u,i^\prime}~\forall~ i\neq i^\prime$. For the sake of simplicity, we will complete the analysis of error probability for the case where $\ell=2$, and then extend the results to a general $\ell$.

\emph{Case $\{\ell=2\}$:} If $\ell=2$, the subsets in \eqref{eq:PairwiseSubsets} can be written as
\begin{align}
\mathcal{J}_{\{1\}} = &\left\{j\left|\mathbf{U}_{j,f,R}=\mathbf{U}_{j_R,f,R}\oplus\left[\begin{array}{c} \underline{e}_{u,1} \\ \underline{0} \end{array}\right]\right.\right\} \nonumber \\
\mathcal{J}_{\{2\}} = &\left\{j\left|\mathbf{U}_{j,f,R}=\mathbf{U}_{j_R,f,R}\oplus\left[\begin{array}{c} \underline{0} \\ \underline{e}_{u,1} \end{array}\right]\right.\right\} \nonumber \\
\mathcal{J}_{\{1,2\}} = &\left\{j\left|\mathbf{U}_{j,f,R}=\mathbf{U}_{j_R,f,R}\oplus\left[\begin{array}{c} \underline{e}_{u,1} \\ \underline{e}_{u,1} \end{array}\right]\right.\right\} \nonumber \\
\mathcal{J}_{\{1\}\{2\}} = &\left\{j\left|\mathbf{U}_{j,f,R}=\mathbf{U}_{j_R,f,R}\oplus\left[\begin{array}{c} \underline{e}_{u,1} \\ \underline{e}_{u,2} \end{array}\right]\right.\right\}. \nonumber
\end{align}
These subsets are disjoint and cover each error event, so
\[\mathcal{J}_{\{1\}}\cup\mathcal{J}_{\{2\}}\cup\mathcal{J}_{\{1,2\}}\cup\mathcal{J}_{\{1\}\{2\}}=\{0,...,2^{2K}-1\}\setminus\{j_R\}.\]
Therefore, the union bound on the probability of error for $\ell=2$ can be written as
\begin{align} \label{eq:PerrSplitL2}
P(Err|j_R) \leq &P(\overline{E}_{j_R}|j_R) + \sum_{j\in\mathcal{J}_{\{1\}}}P(E_j|j_R,j\in\mathcal{J}_{\{1\}}) \nonumber \\
&+\sum_{j\in\mathcal{J}_{\{2\}}}P(E_j|j_R,j\in\mathcal{J}_{\{2\}}) \nonumber \\
&+\sum_{j\in\mathcal{J}_{\{1,2\}}}P(E_j|j_R,j\in\mathcal{J}_{\{1,2\}}) \nonumber \\
&+\sum_{j\in\mathcal{J}_{\{1\}\{2\}}}P(E_j|j_R,j\in\mathcal{J}_{\{1\}\{2\}}).
\end{align}

We define $\underline{e}_{v,i}=\underline{e}_{u,i}\mathbf{G},~i\in\{1,2\}$ as the codeword error vector associated with subset $\mathcal{S}_i$. The subscript $u$ or $v$ is used to differentiate between the message error vector and codeword error vector respectively. Over the ensemble of codes, each $\underline{e}_{v,i}$ is uniformly distributed in $\mathbb{F}_2^N$, and codeword error vectors $\underline{e}_{v,i},\underline{e}_{v,j},~i\neq j$ are pairwise independent and identically distributed. These facts can be shown using steps similar to the proofs of Lemmas 2 and 3.

If $\{j\in\mathcal{J}_{\{1\}}\}$ is given, then we know that
\[\mathbf{X}_{j,f,R}=\mathbf{X}_{j_R,f,R}\oplus\left[\begin{array}{c}\underline{e}_{v,1} \\ \underline{0} \end{array}\right]. \]
By Lemmas 2 and 3, $\underline{v}_{j_R,f,R}^1$ and $\underline{v}_{j,f,R}^1$ are pairwise independent and uniformly distributed on $\mathbb{F}_2^N$. Lemma 2 also tells us that $\underline{v}_{j_R,f,R}^2$ and $\underline{v}_{j,f,R}^2$ are equal and uniformly distributed on $\mathbb{F}_2^N$.

Define $\underline{v}_{R}^2$ as the common value taken by $\underline{v}_{j,f,R}^2=\underline{v}_{j_R,f,R}^2$. Joint AEP provides an asymptotically tight upper bound to each $P(E_j|j_R,j\in\mathcal{J}_{\{1\}})$ if we can show that
\begin{align} \label{eq:J1SameMarginals}
P(\mathbf{X}_{j,f,R},\underline{Y}_R|j_R,j\in\mathcal{J}_{\{1\}}) = P(\mathbf{X}_{j,f,R}|j_R,j\in\mathcal{J}_{\{1\}}) P(\underline{Y}_R|j_R,j\in\mathcal{J}_{\{1\}}).
\end{align}
This is equivalent to showing that
\begin{align} \label{eq:vR2SameMarginals}
P(\mathbf{X}_{j,f,R},\underline{Y}_R|\underline{v}_{R}^2,j\in\mathcal{J}_{\{1\}}) = P(\mathbf{X}_{j,f,R}|\underline{v}_{R}^2,j\in\mathcal{J}_{\{1\}}) P(\underline{Y}_R|\underline{v}_{R}^2,j\in\mathcal{J}_{\{1\}}).
\end{align}
for each value of $\underline{v}_R^2$. Therefore, consider some arbitrary fixed $\underline{v}_R^2$. We can use \eqref{eq:yRDef} and the definition of conditional probability to get
\begin{align}
&P(\mathbf{X}_{j,f,R},\underline{Y}_R|\underline{v}_{R}^2,j\in\mathcal{J}_{\{1\}}) \nonumber \\
&~~~=P(\mathbf{X}_{j,f,R}|\underline{v}_{R}^2,j\in\mathcal{J}_{\{1\}}) P(\underline{Y}_R|\mathbf{X}_{j,f,R},\underline{v}_{R}^2,j\in\mathcal{J}_{\{1\}}) \nonumber \\
&~~~=P(\mathbf{X}_{j,f,R}|\underline{v}_{R}^2,j\in\mathcal{J}_{\{1\}}) P(\underline{Q}_R+\underline{W}_R|\underline{v}_{j,f,R}^1,\underline{v}_{j,f,R}^2,\underline{v}_{R}^2,j\in\mathcal{J}_{\{1\}}).
\end{align}

We see that, conditioned on $\underline{v}_R^2$, $\underline{q}_R$ is a random function of $\underline{v}_{j_R,f,R}^1$,
\[\underline{q}_R = g(\underline{v}_{j_R,f,R}^1;\underline{v}_R^2)\]
which is defined elementwise by Lemma 4. By Lemma 3, $\underline{v}_{j,f,R}^1$ and $\underline{v}_{j_R,f,R}^1$ are pairwise independent, therefore $\underline{v}_{j,f,R}^1$ and $\underline{q}_R$ are independent. Since $\underline{v}_{j,f,R}^2=\underline{v}_R^2$ we have
\[P(Event|\underline{v}_{j,f,R}^2,\underline{v}_R^2,j\in\mathcal{J}_{\{1\}})=P(Event|\underline{v}_R^2,j\in\mathcal{J}_{\{1\}}) \]
for any event. We conclude that
\begin{align}
&P(\underline{Q}_R+\underline{W}_R|\underline{v}_{j,f,R}^1,\underline{v}_{j,f,R}^2,\underline{v}_{R}^2,j\in\mathcal{J}_{\{1\}}) \nonumber \\
&~~~= P(\underline{Q}_R+\underline{W}_R|\underline{v}_{j,f,R}^2,\underline{v}_{R}^2,j\in\mathcal{J}_{\{1\}}) \nonumber \\
&~~~= P(\underline{Q}_R+\underline{W}_R|\underline{v}_{R}^2,j\in\mathcal{J}_{\{1\}}).
\end{align}
This allows us to conclude that \eqref{eq:vR2SameMarginals} holds so we can use \cite[Theorem 15.2.3]{cover2006elements} to get the following bound
\begin{equation}
P(E_j|j_R,j\in\mathcal{J}_{\{1\}}) < 2^{-N(I(Y_R;X_R^1|X_R^2)-3\epsilon)}.
\end{equation}

Similar steps can be used for the case when $j\in\mathcal{J}_{\{2\}}$ to get
\begin{equation}
P(E_j|j_R,j\in\mathcal{J}_{\{2\}}) < 2^{-N(I(Y_R;X_R^2|X_R^1)-3\epsilon)}.
\end{equation}

For the case when $j\in\mathcal{J}_{\{1,2\}}$, we have
\[\mathbf{X}_{j,f,R}=\mathbf{X}_{j_R,f,R}\oplus\left[\begin{array}{c}\underline{e}_{v,1} \\ \underline{e}_{v,1}\end{array}\right]. \]
The most direct way to find a bound for this case is to reassign the address vectors so that this case is similar to the case when $j\in\mathcal{J}_{\{1\}}$. Define a binary matrix $\mathbf{D}_{\{1,2\}}$ given by
\begin{equation}
\mathbf{D}_{\{1,2\}} = \left[\begin{array}{cc} 1 & 0 \\ 1 & 1 \end{array}\right].
\end{equation}
Then define a mapping function $\widetilde{\mathcal{M}}_{f,R}(\cdot)$ by
\begin{equation}
\widetilde{\mathcal{M}}_{f,R}(\underline{x}_{f,R}[n]) = \mathcal{M}_{f,R}(\mathbf{D}_{\{1,2\}}\underline{x}_{f,R}[n]).
\end{equation}
Then define effective codeword matrices $\widetilde{\mathbf{X}}_{j_R,f,R}$ and $\widetilde{\mathbf{X}}_{j,f,R}$ by
\begin{align}
\widetilde{\mathbf{X}}_{j_R,f,R} &= \mathbf{D}_{\{1,2\}} \mathbf{X}_{j_R,f,R} \nonumber \\
\widetilde{\mathbf{X}}_{j,f,R} &= \mathbf{D}_{\{1,2\}} \mathbf{X}_{j,f,R} = \widetilde{\mathbf{X}}_{j_R,f,R}\oplus\left[\begin{array}{c} \underline{e}_{v,1} \\ \underline{0}  \end{array}\right]. \nonumber
\end{align}
This is the same as the case where $j\in\mathcal{J}_{\{1\}}$ if the relay observes the $\underline{y}_R$ corresponding to codeword matricies, $\widetilde{\mathbf{X}}_{j_R,f,R}$ with the mapping function $\widetilde{\mathcal{M}}_{f,R}$. Therefore for the case where $j\in\mathcal{J}_{\{1,2\}}$, we have the bound
\begin{equation}
P(E_j|j_R,j\in\mathcal{J}_{\{1,2\}}) < 2^{-N(I(Y_R;\widetilde{X}_R^1|\widetilde{X}_R^2)-3\epsilon)}
\end{equation}
which can be expressed in terms of the original address variables as
\begin{equation}
P(E_j|j_R,j\in\mathcal{J}_{\{1,2\}}) < 2^{-N(I(Y_R;X_R^1|X_R^1\oplus X_R^2)-3\epsilon)}.
\end{equation}
By the definition of mutual information, we have
\begin{align}
&I(Y_R;X_R^1|X_R^1\oplus X_R^2) \nonumber \\
&~~~= H(Y_R|X_R^1\oplus X_R^2) - H(Y_R|X_R^1,X_R^1\oplus X_R^2) \nonumber \\
&~~~= H(Y_R|X_R^1\oplus X_R^2) - H(Y_R|X_R^1,X_R^2) \nonumber \\
&~~~= I(Y_R;X_R^1,X_R^2|X_R^1\oplus X_R^2). \nonumber
\end{align}
Therefore the bound is equivalent to
\begin{equation}
P(E_j|j_R,j\in\mathcal{J}_{\{1,2\}}) < 2^{-N(I(Y_R;X_R^1,X_R^2|X_R^1\oplus X_R^2)-3\epsilon)}.
\end{equation}

Lastly, for the case when $j\in\mathcal{J}_{\{1\}\{2\}}$, $\mathbf{X}_{j_R,f,R}$ and $\mathbf{X}_{j,f,R}$ are i.i.d. by Lemmas 2 and 3. We can therefore use joint AEP directly to get the bound
\begin{equation}
P(E_j|j_R,j\in\mathcal{J}_{\{1\}\{2\}}) < 2^{-N(I(Y_R;X_R^1,X_R^2)-3\epsilon)}.
\end{equation}

Applying the upper bounds for each index set to \eqref{eq:PerrSplitL2}, we get the following bound
\begin{align}
P(Err|j_R) < &~\epsilon + \sum_{j\in\mathcal{J}_{\{1\}}}2^{-N(I(Y_R;X_R^1|X_R^2)-3\epsilon)} \nonumber \\
&+ \sum_{j\in\mathcal{J}_{\{2\}}}2^{-N(I(Y_R;X_R^2|X_R^1)-3\epsilon)} \nonumber \\
&+ \sum_{j\in\mathcal{J}_{\{1,2\}}}2^{-N(I(Y_R;X_R^1,X_R^2|X_R^1\oplus X_R^2)-3\epsilon)} \nonumber \\
&+ \sum_{j\in\mathcal{J}_{\{1\}\{2\}}}2^{-N(I(Y_R;X_R^1,X_R^2)-3\epsilon)}.
\end{align}
There are $2^{N\mathcal{R}}-1$ elements in the sets $\mathcal{J}_{\{1\}},\mathcal{J}_{\{2\}}$, and $\mathcal{J}_{\{1,2\}}$, and there are fewer than $2^{2N\mathcal{R}}$ elements in the last set $\mathcal{J}_{\{1\}\{2\}}$. Thus the upper bound on the probability of error for this code ensemble can be expressed
\begin{align}
P(Err|j_R) <  &\epsilon + 2^{N(\mathcal{R}-I(Y_R;X_R^1|X_R^2)+3\epsilon)} \nonumber \\
&+ 2^{N(\mathcal{R}-I(Y_R;X_R^2|X_R^1)+3\epsilon)} \nonumber \\
&+ 2^{N(\mathcal{R}-I(Y_R;X_R^1,X_R^2|X_R^1\oplus X_R^2)+3\epsilon)} \nonumber \\
&+ 2^{N(2\mathcal{R}-I(Y_R;X_R^1,X_R^2)+3\epsilon)}.
\end{align}
Each of these terms can be made arbitrarily close to zero by increasing $N$ as long as $\mathcal{R}$ satisfies
\begin{align}
\mathcal{R}<\max(I(Y_R;X_R^1|X_R^2),I(Y_R;X_R^2|X_R^1),I(Y_R;X_R^1,X_R^2|X_R^1\oplus X_R^2),\frac{1}{2}I(Y_R;X_R^1,X_R^2)).
\end{align}
Note that this proof holds for an arbitrary $j_R$ which means that the bound holds independent of the transmitted message.

\emph{Case $\{\ell\geq2\}$:} For a general $\ell$, the proof is very similar. We split \eqref{eq:PerrUnionBound} into the disjoint classes of error events in \eqref{eq:PairwiseSubsets} to get
\begin{align} \label{eq:PerrSplitL}
P(Err|j_R) \leq P(\overline{E}_{j_R}|j_R)+\sum_{p=1}^\ell \sum_{\mathcal{S}_1,...,\mathcal{S}_p} \sum_{j\in\mathcal{J}_{\mathcal{S}_1,...,\mathcal{S}_p}} P(E_j|j_R,j\in\mathcal{J}_{\mathcal{S}_1,...,\mathcal{S}_p}).
\end{align}
Then we find upper bounds on the probability of error for different classes of error events.

First, we consider the case where each $\mathcal{S}_m,~m\in\{1,...,p\}$ contains only its smallest element $t_m$. This first case is analogous to the case where $\{j\in\mathcal{J}_{\{1\}}\}$ for the proof when $\ell=2$. By Lemmas 2 and 3, we have
\begin{align} \label{eq:VjVjrJointDistL}
&P(\underline{V}_{j,f,R}^k=\underline{v}_{1},\underline{V}_{j_R,f,R}^k=\underline{v}_{2}|j\in\mathcal{J}_{\mathcal{S}_1,...,\mathcal{S}_p}) = \begin{cases}
2^{-2N}~&,~k \in \mathcal{S} \\
2^{-N}~&,~\underline{v}_{1}=\underline{v}_{2} ~\textrm{and}~ k\not\in\mathcal{S} \\
0~&,~\underline{v}_{1}\neq\underline{v}_{2} ~\textrm{and}~ k\not\in\mathcal{S}.
\end{cases}
\end{align}
That is if $k\in\mathcal{S}$ then $\underline{v}_{j,f,R}^k$ and $\underline{v}_{j_R,f,R}^k$ are independent and uniformly distributed. If $k\not\in\mathcal{S}$ they are equal and uniformly distributed. Let $\underline{v}_R^k,~k\not\in\mathcal{S}$ be the common value taken by the $k_{th}$ row of $\mathbf{X}_{j,f,R}$ and $\mathbf{X}_{j_R,f,R}$.

The joint AEP gives an asymptotically tight upper bound to $P(E_j|j_R,j\in\mathcal{J}_{\mathcal{S}_1,...,\mathcal{S}_p})$ if we can show that
\begin{align} \label{eq:JointAEPCondition}
P(\mathbf{X}_{j,f,R},\underline{Y}_R|j_R,j\in\mathcal{J}_{\mathcal{S}_1,...,\mathcal{S}_p}) =P(\mathbf{X}_{j,f,R}|j_R,j\in\mathcal{J}_{\mathcal{S}_1,...,\mathcal{S}_p}) P(\underline{Y}_R|j_R,j\in\mathcal{J}_{\mathcal{S}_1,...,\mathcal{S}_p}).
\end{align}
This is equivalent to showing that
\begin{align}
&P(\mathbf{X}_{j,f,R},\underline{Y}_R|\{\underline{v}_R^k~,k\not\in\mathcal{S}\},j\in\mathcal{J}_{\mathcal{S}_1,...,\mathcal{S}_p}) \nonumber \\ &~~~=P(\mathbf{X}_{j,f,R}|\{\underline{v}_R^k~,k\not\in\mathcal{S}\},j\in\mathcal{J}_{\mathcal{S}_1,...,\mathcal{S}_p}) P(\underline{Y}_R|\{\underline{v}_R^k~,k\not\in\mathcal{S}\},j\in\mathcal{J}_{\mathcal{S}_1,...,\mathcal{S}_p})
\end{align}
for each possible set of values $\{\underline{v}_R^k~,k\not\in\mathcal{S}\}$. We can use \eqref{eq:yRDef} and the definition of conditional probability to get
\begin{align}
&P(\mathbf{X}_{j,f,R},\underline{Y}_R|\{\underline{v}_R^k~,k\not\in\mathcal{S}\},j\in\mathcal{J}_{\mathcal{S}_1,...,\mathcal{S}_p}) \nonumber \\
&~~~=P(\mathbf{X}_{j,f,R}|\{\underline{v}_R^k~,k\not\in\mathcal{S}\},j\in\mathcal{J}_{\mathcal{S}_1,...,\mathcal{S}_p}) P(\underline{Y}_R|\mathbf{X}_{j,f,R},\{\underline{v}_R^k~,k\not\in\mathcal{S}\},j\in\mathcal{J}_{\mathcal{S}_1,...,\mathcal{S}_p}) \nonumber \\
&~~~=P(\mathbf{X}_{j,f,R}|\{\underline{v}_R^k~,k\not\in\mathcal{S}\},j\in\mathcal{J}_{\mathcal{S}_1,...,\mathcal{S}_p}) P(\underline{Q}_R+\underline{W}_R|\mathbf{X}_{j,f,R},\{\underline{v}_R^k~,k\not\in\mathcal{S}\},j\in\mathcal{J}_{\mathcal{S}_1,...,\mathcal{S}_p}).
\end{align}
Thus the problem simplifies to showing that
\begin{align} \label{eq:CondIndepL1}
&P(\underline{Q}_R+\underline{W}_R|\mathbf{X}_{j,f,R},\{\underline{v}_R^k~,k\not\in\mathcal{S}\},j\in\mathcal{J}_{\mathcal{S}_1,...,\mathcal{S}_p})
\nonumber \\
&~~~=P(\underline{Q}_R+\underline{W}_R|\{\underline{v}_R^k~,k\not\in\mathcal{S}\},j\in\mathcal{J}_{\mathcal{S}_1,...,\mathcal{S}_p}).
\end{align}
Since we are conditioning on $\{\{\underline{v}_{R}^k,~k\not\in\mathcal{S}\},~j\in\mathcal{J}_{\mathcal{S}_1,...,\mathcal{S}_p}\}$, the values taken by $\underline{v}_{j,f,R}^k~,k\not\in\mathcal{S}$ are already given. Therefore we have
\begin{align}
&P(\underline{Q}_R+\underline{W}_R|\mathbf{X}_{j,f,R},\{\underline{v}_R^k~,k\not\in\mathcal{S}\},j\in\mathcal{J}_{\mathcal{S}_1,...,\mathcal{S}_p})
\nonumber \\
&~~~=P(\underline{Q}_R+\underline{W}_R|\{\underline{v}_{j,f,R}^m~,m\in\mathcal{S}\},\{\underline{v}_R^k~,k\not\in\mathcal{S}\},j\in\mathcal{J}_{\mathcal{S}_1,...,\mathcal{S}_p}). \nonumber
\end{align}
The value taken by $\underline{q}_R$ conditioned on $\{\underline{v}_R^k,~k\not\in\mathcal{S}\}$ is a random function of $\{\underline{v}_{j_R,f,R}^m,~m\in\mathcal{S}\}$,
\[\underline{q}_R = g(\{\underline{v}_{j_R,f,R}^m,~m\in\mathcal{S}\};\{\underline{v}_R^k,~k\not\in\mathcal{S}\})\]
which is defined element wise by Lemma 4. Therefore by the independence of $\{\underline{v}_{j,f,R}^m,~m\in\mathcal{S}\}$ and $\{\underline{v}_{j_R,f,R}^m,~m\in\mathcal{S}\}$ we can conclude that $\underline{q}_R$ is conditionally independent of $\{\underline{v}_{j,f,R}^m,~m\in\mathcal{S}\}$ given $\{\{\underline{v}_R^k,~k\not\in\mathcal{S}\},~j\in\mathcal{J}_{\mathcal{S}_1,...,\mathcal{S}_p}\}$. Therefore since $\underline{w}_R$ is independent of any message, we can conclude that \eqref{eq:CondIndepL1} holds. This is equivalent to \eqref{eq:JointAEPCondition} which allows us to apply joint AEP to get the upper bound
\begin{align} \label{eq:JointAEPL}
P(E_j|j_R,j\in\mathcal{J}_{\mathcal{S}_1,...,\mathcal{S}_p}) \leq 2^{-N (I(Y_R;\{X_R^m,~m\in\mathcal{S}\}|\{X_R^k,~k\not\in\mathcal{S}\})-3\epsilon)}.
\end{align}

To extend this result to the case where each $\mathcal{S}_1,...,\mathcal{S}_p$ can contain multiple elements, we make this problem look like the first case. Define a matrix $\mathbf{D}_{\mathcal{S}_1,...,\mathcal{S}_p}$ whose $m_{th}$ column $\underline{d}_m$ is given by
\begin{align}
\underline{d}_{t_k}[n] &= \begin{cases} 1~,~ n\in\mathcal{S}_k \\
0~,~ n\not\in\mathcal{S}_k
\end{cases} \forall ~ k=1,...,p \nonumber \\
\underline{d}_m[n] &= \begin{cases} 1~,~ n=m \\
0~,~ n\neq m
\end{cases} \forall ~ m\not\in\mathcal{T}.
\end{align}

For example, if $\ell=6$, $\mathcal{S}_1=\{2,4,5\}$, and $\mathcal{S}_2=\{3,6\}$ we have
\[
\mathbf{D}_{\{2,4,5\},\{3,6\}} = \left[\begin{array}{cccccc} 1&0&0&0&0&0 \\ 0&1&0&0&0&0 \\ 0&0&1&0&0&0 \\ 0&1&0&1&0&0 \\ 0&1&0&0&1&0 \\ 0&0&1&0&0&1 \end{array} \right].
\]
Define effective codeword matrices $\widetilde{\mathbf{X}}_{j,f,R}$ and $\widetilde{\mathbf{X}}_{j_R,f,R}$ by
\begin{align} \label{eq:XTildeDef}
\widetilde{\mathbf{X}}_{j,f,R} &= \mathbf{D}_{\mathcal{S}_1,...,\mathcal{S}_p} \mathbf{X}_{j,f,R} \nonumber \\
\widetilde{\mathbf{X}}_{j_R,f,R} &= \mathbf{D}_{\mathcal{S}_1,...,\mathcal{S}_p} \mathbf{X}_{j_R,f,R}.
\end{align}

Then the $k_{th}$ row $\widetilde{\underline{v}}_{j,f,R}^k$ of $\widetilde{\mathbf{X}}_{j,f,R}$ is given by
\begin{align}
\widetilde{\underline{v}}_{j,f,R}^k = \begin{cases}
\widetilde{\underline{v}}_{j_R,f,R}^k&,~ k\not\in\mathcal{T} \\
\widetilde{\underline{v}}_{j_R,f,R}^k\oplus\underline{e}_{v,m}&,~k=t_m,~ m=1,...,p
\end{cases}
\end{align}
for some set of pairwise independent error vectors $\underline{e}_{v,1},...,\underline{e}_{v,p}\in\mathbb{F}_2^N\setminus\{\underline{0}\}$.

For the $\ell=6$ example, this means that
\[
\widetilde{\mathbf{X}}_{j,f,R} = \widetilde{\mathbf{X}}_{j_R,f,R}\oplus \left[\begin{array}{c} \underline{0} \\ \underline{e}_{1,v} \\ \underline{e}_{2,v} \\ \underline{0} \\ \underline{0} \\ \underline{0} \end{array} \right].
\]
This is the same as the case where each $\mathcal{S}_1,...,\mathcal{S}_p$ contains only one element. Thus, we can apply the bound in \eqref{eq:JointAEPL} to get
\begin{align} \label{eq:JointAEPLtilde}
P(E_j|j_R,j\in\mathcal{J}_{\mathcal{S}_1,...,\mathcal{S}_p}) \leq 2^{-N(I(Y_R;\{\widetilde{X}_R^k,~k\in\mathcal{T}\}|\{\widetilde{X}_R^k,~k\not\in\mathcal{T}\})-3\epsilon)}.
\end{align}

The only step that remains is to show that the mutual information in \eqref{eq:JointAEPLtilde} can be expressed as
\begin{align} \label{eq:IyxAEPLequiv}
&I(Y_R;\{\widetilde{X}_R^k,~k\in\mathcal{T}\}|\{\widetilde{X}_R^k,~k\not\in\mathcal{T}\}) \nonumber \\
&~~~= I(Y_R;\{X_R^k|k\in\mathcal{S}\}|\{X_R^k|k\not\in\mathcal{S}\},\{X_R^k\oplus Z_i|k\in\mathcal{S}_i\}~\forall ~i=1,...,p)
\end{align}
where $\mathcal{S}=\bigcup_{i=1}^p\mathcal{S}_i$, and each $Z_i$ is an auxiliary random variable which is Bernoulli distributed with parameter $\frac{1}{2}$. By \eqref{eq:XTildeDef}, we have
\begin{align}
\widetilde{x}_R^k = \begin{cases}
x_R^k ~&,~ k\in\overline{\mathcal{S}}\cup\mathcal{T} \\
x_R^k\oplus x_R^{t_m} ~&,~ k\in\mathcal{S}_m\setminus\{t_m\}.
\end{cases}
\end{align}
We therefore have
\begin{align}
\{\widetilde{x}_R^k|k\in\mathcal{T}\} &\Leftrightarrow \{x_R^k|k\in\mathcal{T}\} \nonumber \\
\{\widetilde{x}_R^k|k\not\in\mathcal{T}\} &\Leftrightarrow \{x_R^k|k\not\in\mathcal{S}\}\cup \bigcup_{m=1}^{p} \{x_R^k\oplus x_R^{t_m}|k\in\mathcal{S}_m\setminus\{t_m\}\}. \nonumber
\end{align}
The mutual information can therefore be expressed
\begin{align}
&I(Y_R;\{\widetilde{X}_R^k|k\in\mathcal{T}\}|\{\widetilde{X}_R^k|k\not\in\mathcal{T}\}) \nonumber \\
&~~~=I(Y_R;\{X_R^k|k\in\mathcal{T}\}|\{X_R^k|k\not\in\mathcal{S}\}\cup \bigcup_{m=1}^{p} \{X_R^k\oplus X_R^{t_m}|k\in\mathcal{S}_m\setminus\{t_m\}\}) \nonumber \\
&~~~=H(Y_R|\{X_R^k|k\not\in\mathcal{S}\}\cup \bigcup_{m=1}^{p} \{X_R^k\oplus X_R^{t_m}|k\in\mathcal{S}_m\setminus\{t_m\}\})-H(Y_R|X_R^1,...,X_R^\ell). \nonumber
\end{align}
The last equality follows because if we know $x_R^{t_m}$ and $x_R^k\oplus x_R^{t_m}$ then we know both $x_R^{t_m}$ and $x_R^k$. Which tells us that knowing $\{x_R^k|k\in\mathcal{T}\}\cup\bigcup_{m=1}^{p} \{x_R^k\oplus x_R^{t_m}|k\in\mathcal{S}_m\setminus\{t_m\}\}$ is equivalent to knowing $\{x_R^k|k\in\mathcal{S}\}$.

It can be shown for each $\mathcal{S}_m,~ m\in\{1,...,p\}$ that
\begin{equation}
\{x_R^k\oplus x_R^{t_m}|k\in\mathcal{S}_m\setminus\{t_m\}\} \Leftrightarrow \{x_R^k\oplus z_m|k\in\mathcal{S}_m\}.
\end{equation}
For example, if we consider our $\ell=6$ case, we have $\mathcal{S}_1=\{2,4,5\}$. If we know that
\[(x_R^4 \oplus x_R^2,x_R^5 \oplus x_R^2)=(a,b)~,~a,b\in\{0,1\}\]
then we have
\[(x_R^2,x_R^4,x_R^5)\in\{(0,a,b),(1, \overline{a},\overline{b})\}\]
which is equivalent to knowing
\[(x_R^2\oplus z_1,x_R^4\oplus z_1,x_R^5\oplus z_1).\]
We therefore have
\begin{align}
&I(Y_R;\{\widetilde{X}_R^k|k\in\mathcal{T}\}|\{\widetilde{X}_R^k|k\not\in\mathcal{T}\}) \nonumber \\
&~~~=H(Y_R|\{X_R^k|k\not\in\mathcal{S}\}\cup \bigcup_{m=1}^{p} \{X_R^k\oplus X_R^{t_m}|k\in\mathcal{S}_m\setminus\{t_m\}\})-H(Y_R|X_R^1,...,X_R^\ell) \nonumber \\
&~~~=H(Y_R|\{X_R^k|k\not\in\mathcal{S}\}\cup \bigcup_{m=1}^{p} \{X_R^k\oplus Z_m|k\in\mathcal{S}_m\})-H(Y_R|X_R^1,...,X_R^\ell) \nonumber \\
&~~~=I(Y_R;\{X_R^k|k\in\mathcal{S}\}|\{X_R^k|k\not\in\mathcal{S}\},\{X_R^k\oplus Z_i|k\in\mathcal{S}_i\}~\forall ~i=1,...,p).
\end{align}
This is the same as \eqref{eq:IyxAEPLequiv}, which allows us to restate the bound in \eqref{eq:JointAEPLtilde} as
\begin{align}
&P(E_j|j_R,j\in\mathcal{J}_{\mathcal{S}_1,...,\mathcal{S}_p}) \nonumber \\
&~~~\leq 2^{-N (I(Y_R;\{X_R^k|k\in\mathcal{S}\}|\{X_R^k|k\not\in\mathcal{S}\},\{X_R^k\oplus Z_i|k\in\mathcal{S}_i\}~\forall ~i=1,...,p)-3\epsilon)} \nonumber \\
&~~~\triangleq 2^{-N (I(Y_R;\mathcal{S}_1,...,\mathcal{S}_p)-3\epsilon)}.
\end{align}
The last step defines a mutual information $I(Y_R;\mathcal{S}_1,...,\mathcal{S}_p)$. This slight abuse of notation simplifies the last few steps of the proof.

Plugging this into \eqref{eq:PerrSplitL}, we have
\begin{align}
&P(Err|j_R) \leq P(\overline{E}_{j_R}|j_R)+\sum_{p=1}^\ell \sum_{\mathcal{S}_1,...,\mathcal{S}_p} \sum_{j\in\mathcal{J}_{\mathcal{S}_1,...,\mathcal{S}_p}} 2^{-N (I(Y_R;\mathcal{S}_1,...,\mathcal{S}_p)-3\epsilon)}.
\end{align}
For each possible $\mathcal{S}_1,...,\mathcal{S}_p$ we have
\begin{equation}
|\mathcal{J}_{\mathcal{S}_1,...,\mathcal{S}_p}|=(2^{N\mathcal{R}}-1)(2^{N\mathcal{R}}-2)...(2^{N\mathcal{R}}-p)<2^{N\mathcal{R}p}. \nonumber
\end{equation}
Therefore we have
\begin{align}
P(Err|j_R) &\leq \epsilon + \sum_{p=1}^\ell \sum_{\mathcal{S}_1,...,\mathcal{S}_p} 2^{N\mathcal{R}p} 2^{-N (I(Y_R;\mathcal{S}_1,...,\mathcal{S}_p)-3\epsilon)} \nonumber \\
&\leq \epsilon + \sum_{p=1}^\ell \sum_{\mathcal{S}_1,...,\mathcal{S}_p}2^{N (\mathcal{R}p-I(Y_R;\mathcal{S}_1,...,\mathcal{S}_p)+3\epsilon)}. \nonumber
\end{align}
This bound approaches zero as long as
\begin{align}
\mathcal{R}< \underset{\mathcal{S},\overline{\mathcal{S}},\mathcal{S}_1,...,\mathcal{S}_p}{min} ~ \frac{1}{p} I(Y_R;\{X_R^k|k\in\mathcal{S}\}|\{X_R^k|k\in\overline{\mathcal{S}}\}, \{X_R^k\oplus Z_i|k\in\mathcal{S}_i\} ~\forall~ i\in\{1,...,p\}).
\end{align}
This completes the proof.

\bibliographystyle{ieeetr}
\bibliography{lattice}

\end{document}